\documentclass[pdflatex,sn-mathphys-num]{sn-jnl}

\usepackage{graphicx}%
\usepackage{multirow}%
\usepackage{amsmath,amssymb,amsfonts}%
\usepackage{amsthm}%
\usepackage{mathrsfs}%
\usepackage[title]{appendix}%
\usepackage{xcolor}%
\usepackage{textcomp}%
\usepackage{manyfoot}%
\usepackage{booktabs}%
\usepackage{algorithm}%
\usepackage{algorithmicx}%
\usepackage{algpseudocode}%
\usepackage{listings}%
\usepackage{bm}
\usepackage{geometry}
\theoremstyle{thmstyleone}%
\theoremstyle{thmstyletwo}%

\theoremstyle{thmstylethree}%

\UseRawInputEncoding

\begin{document}

\title[Article Title]{A Universal Spin-Orbit-Coupled Hamiltonian Model for Accelerated Quantum Material Discovery}


\author[1]{\fnm{Yang} \sur{Zhong}}
\equalcont{These authors contributed equally to this work.}

\author[1]{\fnm{Rui} \sur{Wang}}
\equalcont{These authors contributed equally to this work.}

\author[1]{\fnm{Xingao} \sur{Gong}}

\author*[1]{\fnm{Hongjun} \sur{Xiang}}\email{hxiang@fudan.edu.cn}

\affil[1]
{\orgdiv{Key Laboratory of Computational Physical Sciences (Ministry of Education)}, 
\orgname{Institute of Computational Physical Sciences, State Key Laboratory of Surface Physics}, 
\orgname{and Department of Physics}, 
\orgname{Fudan University},
\orgaddress{\city{Shanghai}, \postcode{200433},  \country{China}}}


\abstract{The accurate modeling of spin-orbit coupling (SOC) effects in diverse complex systems remains a significant challenge due to the high computational demands of density functional theory (DFT) and the limited transferability of existing machine-learning frameworks. This study addresses these limitations by introducing Uni-HamGNN, a universal SOC Hamiltonian graph neural network that is applicable across the periodic table. By decomposing the SOC Hamiltonian into spin-independent and SOC correction terms, our approach preserves SU(2) symmetry while significantly reducing parameter requirements. Based on this decomposition, we propose a delta-learning strategy to separately fit the two components, thereby addressing the training difficulties caused by magnitude discrepancies between them and enabling efficient training. The model achieves remarkable accuracy (mean absolute error of 0.0025 meV for the SOC-related component) and demonstrates broad applicability through high-throughput screening of the GNoME dataset for topological insulators, as well as precise predictions for 2D valleytronic materials and transition metal dichalcogenide (TMD) heterostructures. This breakthrough eliminates the need for system-specific retraining and costly SOC-DFT calculations, paving the way for rapid discovery of quantum materials.}


\maketitle

\section{Introduction}\label{sec1}

The spin-orbit coupling (SOC) effect has emerged as a cornerstone of spintronics, enabling all-electrical control of spin states without requiring external magnetic fields\cite{bib1, bib2}. This interaction underpins transformative quantum phenomena and devices, such as the spin Hall effect (SHE)\cite{bib3}, topological insulators (TIs)\cite{bib4, bib5, bib6, bib7}, valleytronics\cite{bib8, bib9}, and the spin field-effect transistors (SFT)\cite{bib10}. SOC-driven mechanisms offer unprecedented opportunities for low-dissipation spin manipulation and topological quantum materials design, making precise calculation of SOC electronic structures indispensable for advancing next-generation electronic devices\cite{bib10, bib11}. Despite its fundamental importance, accurately modeling SOC effects in large systems remains computationally prohibitive for DFT, creating a bottleneck in the design of next-generation quantum materials.

Recent advancements in data-driven Hamiltonian parameterization may provide new solutions to the long-standing challenges\cite{bib12,bib13,bib14, bib15,bib16,bib17,bib18,bib19}. By constructing direct mappings from atomic environments to Hamiltonian matrices, these models bypass the computationally expensive self-consistent field (SCF) iterations in DFT, enabling efficient predictions of electronic structures. Recently, a spinless universal Hamiltonian model applicable to arbitrary element combinations has been reported\cite{bib20}. This model successfully addresses the challenges of extrapolating to new material systems and eliminates the need for resource-intensive re-training for each new composition. However, the development of a universal SOC Hamiltonian model still encounters three major challenges that require careful attention and innovative solutions: First, generating SOC Hamiltonian training sets involves computationally intensive DFT calculations that include SOC, which are approximately eight times slower than their non-SOC counterparts. Second, incorporating SOC transforms Hamiltonians into complex-valued matrices with quadrupled dimensions, dramatically expanding the parameter space. Third, the two-orders-of-magnitude disparity between SOC and non-SOC matrix elements can lead to training instability, as conventional neural networks struggle to discern the subtle contributions of SOC in the presence of dominant spin-independent terms.

Our work addresses these challenges through a physics-informed decomposition strategy that separates SOC Hamiltonians into spin-independent components and symmetry-preserving correction terms. This method not only reduces the number of required parameters but also rigorously maintains SU(2) symmetry, thereby enabling a delta-learning framework capable of independently optimizing magnitude-disparate terms in separate channels. Building upon this foundation, we integrate the dual-channel framework with HamGNN-V2, an enhanced graph neural network architecture that extends HamGNN by incorporating optimized convolutional layers. The resulting universal SOC Hamiltonian model, termed Uni-HamGNN, was trained using a resource-efficient dataset containing only 10,000 SOC matrices supplemented by 40,000 computationally economical non-SOC matrices. Using Uni-HamGNN, we conducted high-throughput screening of topological insulators within the GMoME dataset containing 300,000 inorganic crystal structures, showcasing the model's efficiency and accuracy in identifying topological materials. Further validation on two-dimensional materials demonstrates that Uni-HamGNN exhibits high accuracy in predicting SOC electronic structures in valleytronic materials and bilayer transition metal dichalcogenide (TMD) heterostructures, illustrating the model's broad applicability and high transferability across different dimensional material systems. These successful applications indicate that Uni-HamGNN holds promise as a revolutionary tool for quantum material design and property research, advancing the cutting-edge study of condensed matter physics and materials science to new heights.

\section{Results}\label{sec2}
\subsection{Physics-Informed Framework for Parameterizing the SOC Hamiltonian}\label{subsec2}

The electronic Hamiltonian matrix comprises a substantial number of elements that encode interactions and energy information between orbitals of distinct atoms within a system. For the spinless Hamiltonian $H_{\mathrm{0}}$, its parameterization typically leverages its equivariant constraint under the $\mathrm{SO}(3)$ rotation operation, representing it as a set of learnable irreducible spherical tensors (ISTs)\cite{bib17,bib21}. However, introducing SOC effects fundamentally alters the Hamiltonian’s complexity. The SOC Hamiltonian transforms into a complex-valued matrix of dimensions $2N_{\mathrm{orb}} \times 2N_{\mathrm{orb}}$, where $N_{\mathrm{orb}}$ denotes the number of localized orbitals. This expansion---driven by the inclusion of spin degrees of freedom---quadruples the matrix dimensions relative to the spinless case, while the transition to complex matrices further doubles the number of elements. A conservative estimate suggests that adopting an analogous parameterization approach for the SOC Hamiltonian would require approximately \emph{eight times} as many parameters as the spinless case. Additionally, the SOC Hamiltonian must simultaneously satisfy rotational symmetry constraints for both orbital angular momentum ($l$) and spin angular momentum ($s$), leading to the $\mathrm{SU}(2)$ transformation rule in the spin-orbit basis:

\begin{equation}
H_{\mathrm{SOC}} = D^{(l)}(Q) \otimes D^{(1/2)}(Q) \cdot H_{\mathrm{SOC}} \cdot \left[D^{(l)}(Q) \otimes D^{(1/2)}(Q)\right]^\dag 
\tag{1}
\end{equation}

\noindent Here, $D^{(l)}(Q)$ and $D^{(1/2)}(Q)$ denote the Wigner rotation matrices associated with orbital and spin angular momentum\cite{bib22}, respectively, under the rotation $Q \in \mathrm{SO}(3)$. The combination of high-dimensional parameterization and $\mathrm{SU}(2)$ symmetry constraints presents a formidable challenge for constructing a universal SOC Hamiltonian model applicable across the periodic table.

By shifting our focus to the physical interpretation of the SOC effect, we can identify more straightforward parameterization approaches for the SOC Hamiltonian. The classical interpretation attributes SOC to the interaction between the electron's spin magnetic moment $\vec{\mu}$ and the effective magnetic field $\vec{B}$ from its orbital motion, yielding an energy correction $-\vec{\mu} \cdot \vec{B}$. This manifests quantum mechanically as a Hamiltonian term $H_{\text{soc}} \sim \vec{L} \cdot \vec{\sigma}$, where $\vec{L}$ is the orbital angular momentum operator and $\vec{\sigma}$ the Pauli spin vector. The total Hamiltonian thus decomposes into spin-independent ($H_0$) and SOC-dependent ($H_{\text{soc}}$) terms:
\begin{equation}
H_0 \otimes I_2 + H_{\text{soc}} 
\tag{2}
\end{equation}
\noindent Here, $I_2$ is the $2 \times 2$ identity matrix. The SOC term is explicitly parameterized as\cite{bib23}:

\begin{equation}
H_{\text{soc}} = \xi \widehat{\vec{L}} \cdot \widehat{\vec{\sigma}} = \xi \left( \hat{L}_x \hat{\sigma}_x + \hat{L}_y \hat{\sigma}_y + \hat{L}_z \hat{\sigma}_z \right) 
\tag{3}
\end{equation}

\begin{figure}[t]
\centering
\includegraphics[width=0.9\textwidth]{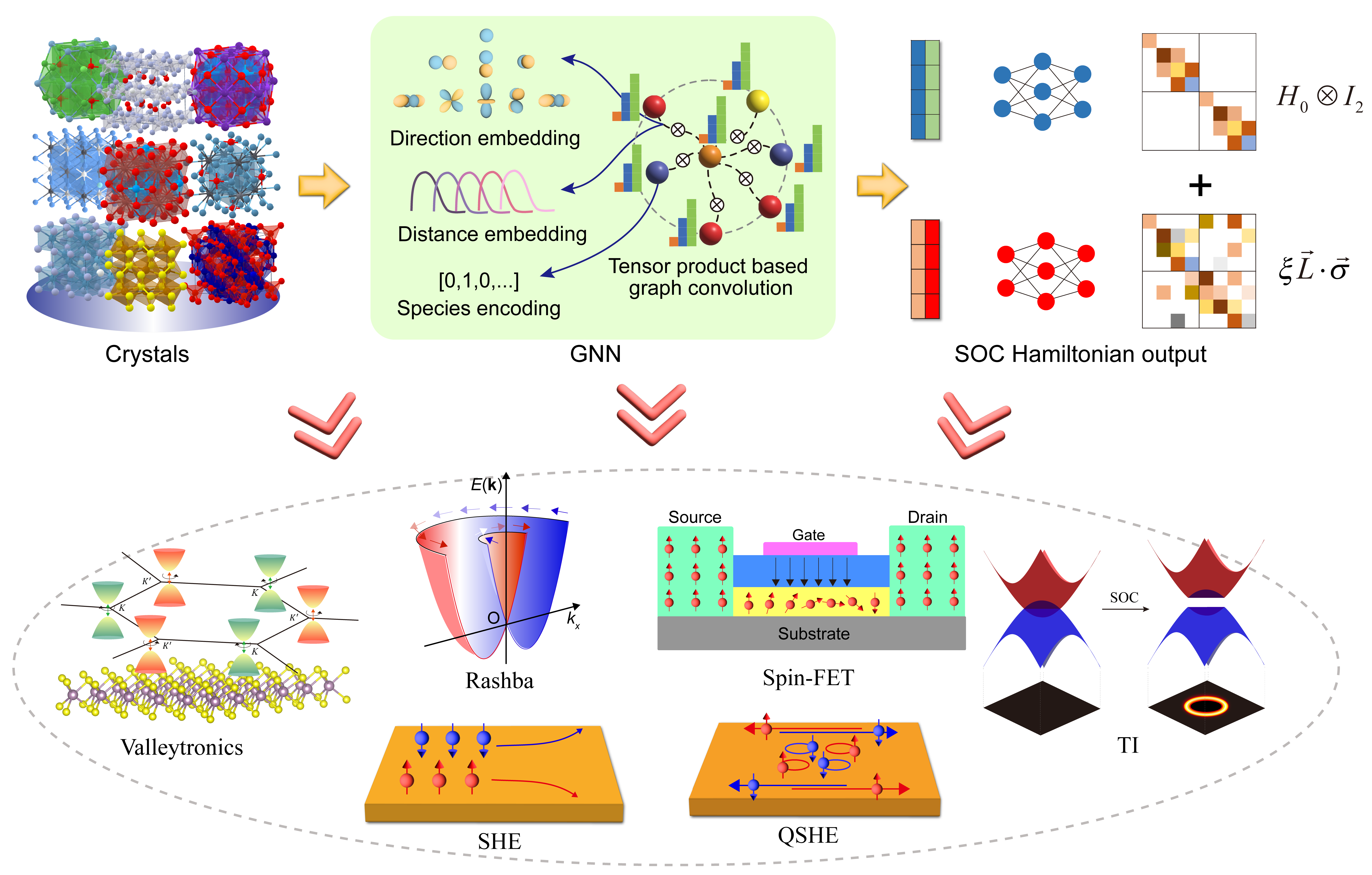}
\caption{\textbf{The framework for parameterizing spin-orbit coupling (SOC) Hamiltonians using graph neural networks (GNNs).} The GNN effectively encodes atomic species, bond orientations, and bond lengths within crystal structures, performing graph convolution through equivariant tensor product operations based on these features. At the output stage, the model constructs two components based on the learned features: the spin-independent Hamiltonian matrix $H_0$ and the spin-orbit coupling term $\xi \widehat{\vec{L}} \cdot \widehat{\vec{\sigma}}$, where $\xi$ denotes the SOC strength factor, $\widehat{\vec{L}}$ represents the orbital angular momentum operator, and $\widehat{\vec{\sigma}}$ corresponds to the Pauli matrices. The complete SOC Hamiltonian matrix, obtained by summing these components, describes electronic structures under SOC effects. This framework enables systematic investigation of diverse SOC-related quantum phenomena, including valley degree of freedom manipulation in valleytronics, spin current mechanisms in the spin Hall effect (SHE) and quantum spin Hall effect (QSHE), unique spin textures induced by Rashba SOC, operational mechanisms of spin field-effect transistors (Spin-FETs), and distinctive physical properties of topological insulator (TI) surface states. These applications rely on precise characterization and modulation of SOC effects in materials, and the framework offers an efficient and accurate modeling approach to achieve these goals.}\label{fig1}
\end{figure}

\noindent where $\xi$ quantifies the SOC strength. Note that the non-SOC part, $H_0$, is always real, whereas $H_{\text{soc}}$ contains both real and imaginary components. This formulation not only clarifies the spin-orbit coupling mechanism but also enables a tractable parameterization in the atomic orbital basis $\left\{ \phi_i, \phi_j \right\}$:

\begin{equation}
H_{ij}^{{s_i}{s_j}} = \begin{cases} 
H_{ij}^0 + \frac{1}{2} \xi_{ij} \left\langle \phi_i \right| \hat{L}_z \left| \phi_j \right\rangle \quad \quad \quad \quad \quad \quad \quad s_i = \uparrow, s_j = \uparrow \\
\frac{1}{2} \xi_{ij} \left( \left\langle \phi_i \right| \hat{L}_x \left| \phi_j \right\rangle - i \left\langle \phi_i \right| \hat{L}_y \left| \phi_j \right\rangle \right)  \quad \quad s_i = \uparrow, s_j = \downarrow \\
\frac{1}{2} \xi_{ij} \left( \left\langle \phi_i \right| \hat{L}_x \left| \phi_j \right\rangle + i \left\langle \phi_i \right| \hat{L}_y \left| \phi_j \right\rangle \right) \quad\quad s_i = \downarrow, s_j = \uparrow \\
H_{ij}^0 - \frac{1}{2} \xi_{ij} \left\langle \phi_i \right| \hat{L}_z \left| \phi_j \right\rangle \quad \quad \quad \quad \quad \quad \quad s_i = \downarrow, s_j = \downarrow 
\end{cases}
\tag{4}
\end{equation}

The parameterization in Eq. (4) significantly reduces the modeling burden by depending on two key components: analytically computable orbital angular momentum matrix elements $\langle \phi_i | \hat{L}_\alpha | \phi_j \rangle$ ($\alpha = x, y, z$) and learnable coefficients $\xi_{ij}$. Our framework systematically incorporates both on-site ($j = i$) and off-site ($j \neq i$) SOC contributions into the Hamiltonian matrix, ensuring robust performance across diverse material systems---including those dominated by heavy elements. This approach contrasts with the empirical tight-binding framework, where SOC is conventionally treated under the on-site approximation, considering only the one-center term $H_{\mathrm{SOC}}^{ii}$\cite{bib23, RN2214}. While this simplification suffices for systems with moderate SOC effects, it becomes inadequate for heavy-element systems, such as those involving 5$d$ transition metals, where two-center SOC terms play a critical role\cite{RN2215, RN2216}. By explicitly including these contributions, our model captures the full complexity of spin-orbit interactions in materials with strong relativistic effects.

Building on this physics-grounded parameterization, we develop a universal SOC Hamiltonian parameterization framework using graph neural networks (GNNs), as illustrated in Fig. 1. The GNN architecture encodes atomic species, bond orientations, and interatomic distances, mapping them through tensor-product convolutions to both the spin-independent Hamiltonian $H_{ij}^0$ and SOC strength $\xi_{ij}$. Training across diverse crystal structures endows the model with robust generalization capabilities, allowing for accurate SOC Hamiltonian predictions even in previously unseen materials and thereby facilitating efficient modeling of SOC-related quantum effects across a variety of complex systems.

\begin{figure}[t]
\centering
\includegraphics[width=0.9\textwidth]{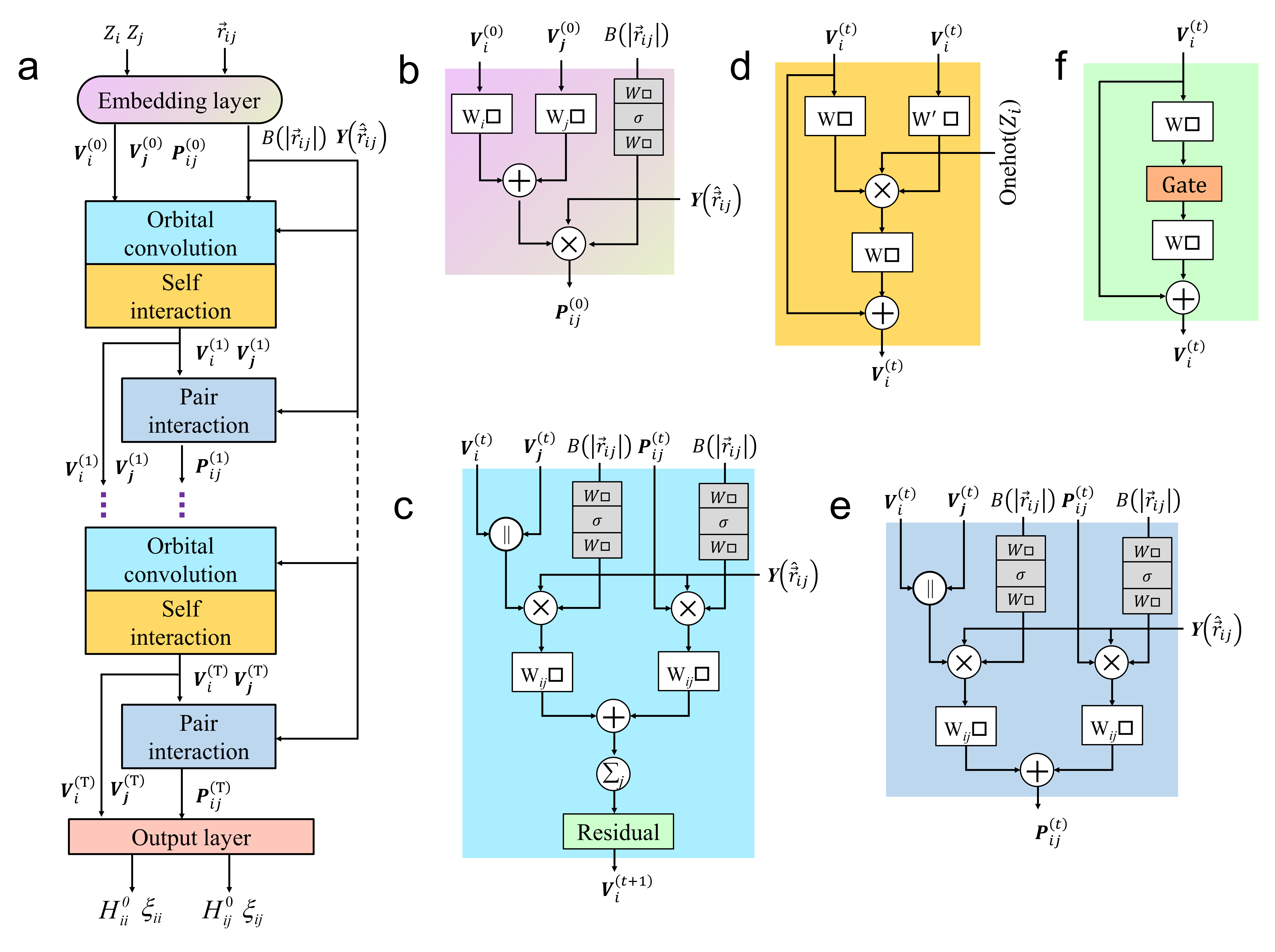}
\caption{\textbf{HamGNN-V2 architecture and the illustration of its subnetworks.} \textbf{a} Network architecture: Initial atom features $V_i^{(0)}$ are one-hot encoded by element. Bond lengths are embedded via Bessel functions $B\left(\left\|\bm{r}_{ij}\right\|\right)$, and orientations via spherical harmonics $Y\left({\hat{\bm{r}}}_{ij}\right)$. The initial edge feature vector $P_{ij}^{(0)}$ is constructed by combining the features of terminal atoms, bond length, and directional information (see panel \textbf{b}). The network recursively updates atomic features $V_i^{(t)}$ and edge features $P_{ij}^{(t)}$ through $T$ layers of orbital convolution, self-interaction, and pair interaction operations. The output layer constructs the non-SOC Hamiltonian matrix $H_0$ and the SOC strength factor $\xi$ from the final features. \textbf{b} Edge initialization: $P_{ij}^{(0)}$ combines one-hot atom encodings and spherical harmonics of the bond through tensor products. \textbf{c} Orbital convolution: Updates atom features via message-passing, using concatenated atomic $V_i^{(t)}\parallel V_j^{(t)}$ and edge features $P_{ij}^{(t)}$ with tensor products to capture atomic orbital characteristics under different rotational orders. \textbf{d} Self-interaction: MACE self-interaction layer models higher-order interactions between atomic orbitals across rotational orders $l$. \textbf{e} Pair interaction: Updates $P_{ij}^{(t)}$ via tensor products of $V_i^{(t)}\parallel V_j^{(t)}$ and edge features $P_{ij}^{(t)}$, integrating local chemical environments. \textbf{f} Residual layers: Apply nonlinear transformations to atomic features with shortcut connections to enhance representational capacity and training stability.}\label{fig2}
\end{figure}

To implement this parameterization framework, we develop HamGNN-V2---an enhanced version of the original HamGNN\cite{bib17} architecture that recently achieved a spin-less universal Hamiltonian model\cite{bib20}. As illustrated in Fig. 2a, the architecture of HamGNN-V2 features a hierarchical graph neural network that processes crystal structures through a sequence of orbital convolution, self-interaction, and pair interaction layers. This neural network inputs a graph representation of a crystal, with each atom represented as a node and each chemical bond as an edge linking two nodes. An initial one-hot vector based on elemental type is constructed for each atom, noted as $V_i^{(0)}$. Bond geometric properties, embedded using a set of Bessel functions, yield the bond-length embedding vector $B\left(|\bm{r}_{ij}|\right)$, while spherical harmonics $Y\left({\hat{\bm{r}}}_{ij}\right)$ capture bond directionality. By integrating initial atomic features with bond-length and directional embeddings, the initial edge feature vector $P_{ij}^{(0)}$ is formed (see Fig. 2b).

Following input feature initialization, the network employs tensor-product-based convolution operations and a message-passing mechanism to update atomic and bond features iteratively (Fig. 2c). After each convolutional layer, the atomic and bond features are refined to $V_i^{(t)}$ and $P_{ij}^{(t)}$ respectively, where $t$ represents the convolutional layer's depth. The incorporation of self-interaction layers (Fig. 2d) from MACE\cite{bib25} further enhances HamGNN-V2, specifically designed to capture interactions between orbital features of varying rotational orders $l$, thus providing high-order features for the orbital convolution layers. The network recursively updates atomic features and edge features through $T$ layers of orbital convolution layers, self-interaction layers, and pair interaction layers (Fig. 2e). Within these layers, we utilize an optimized tensor product operation (detailed in the Methods section), cutting down the required network parameters to $\frac{1}{N_p} + \frac{1}{C_x}$ relative to the traditional tensor product operation, where $N_p$ and $C_x$ stand for the respective total tensor-product paths and feature channels. This innovation curtails computational resource demand while preserving the model's expressive power. Ultimately, the output layer maps these two features to the non-SOC Hamiltonian matrix $H_0$ and the SOC strength factor $\xi$, thereby constructing the complete SOC Hamiltonian matrix.

\subsection{Delta Learning of SOC Hamiltonian}\label{subsec2}
After incorporating the spin degree of freedom, DFT calculations with SOC become substantially more computationally intensive compared to non-SOC DFT calculations. Consequently, constructing a training dataset for SOC Hamiltonians represents a highly demanding computational task. An additional challenge arises from the disparity in magnitude between the non-SOC and SOC terms; the non-SOC term often reaches tens of $\mathrm{eV}$, while the SOC term usually amounts to only a few tenths of an $\mathrm{eV}$. However, despite its smaller magnitude, the SOC term has a profound impact on the electronic structure and material properties, underscoring the difficulty of accurately fitting both terms given their significant amplitude difference.

\begin{figure}[t]
\centering
\includegraphics[width=0.9\textwidth]{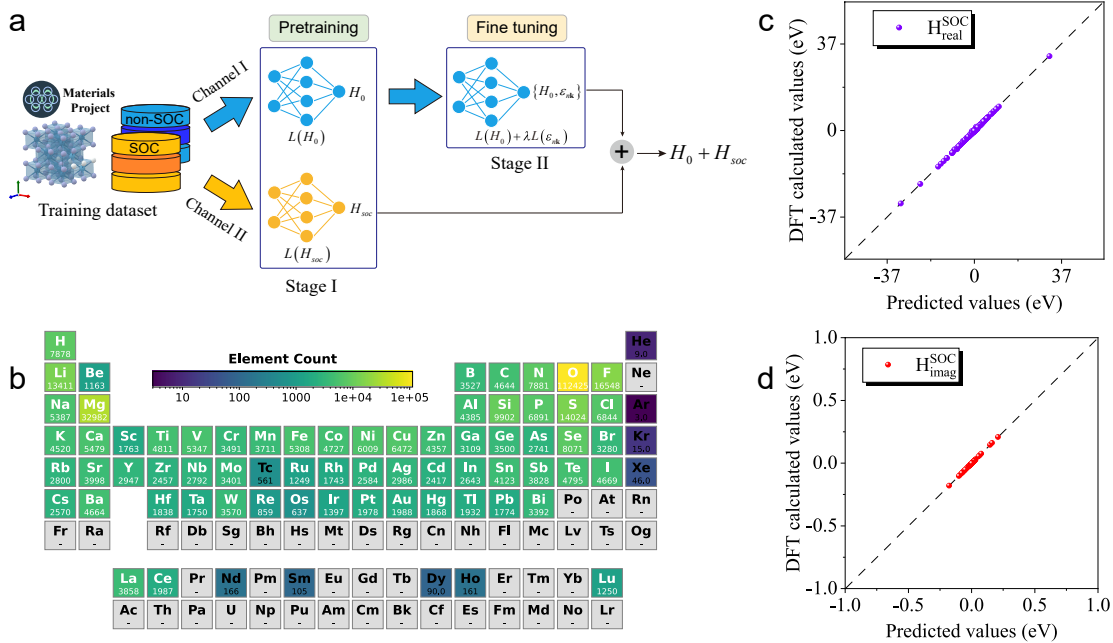}
\caption{\textbf{Overview of the dataset and training framework.} \textbf{a} Schematic of the delta-learning framework for the Universal SOC Hamiltonian Model (Uni-HamGNN). \textbf{b} Element count in the whole training dataset. \textbf{c} Comparison between Uni-HamGNN-predicted real components of SOC Hamiltonians ($H_{\mathrm{real}}^{\mathrm{SOC}}$) and density functional theory (DFT)-computed values on the test set. \textbf{d} Comparison between Uni-HamGNN-predicted imaginary components of SOC Hamiltonians ($H_{\mathrm{imag}}^{\mathrm{SOC}}$) and DFT-computed values on the test set.
}\label{fig3}
\end{figure}

To address this issue, we propose a delta-learning method to train HamGNN-V2 model by separately learning the two components in distinct channels, as illustrated in Fig. 3a. This allows us to utilize a large training set of non-SOC Hamiltonians, which are generated at a lower computational cost, to fit $H_0$, while requiring only a small subset of the imaginary parts of SOC Hamiltonian matrices to train the parameter $\xi$. This approach effectively prevents the loss value associated with the SOC term from being overshadowed or lost during the simultaneous training of all terms. In our study, we employed a dataset comprising non-SOC Hamiltonians derived from 44,000 structures in the Materials Project\cite{bib26} database to train $H_0$. Meanwhile, the imaginary parts of SOC Hamiltonians extracted from approximately 10,000 structures were utilized as training targets for fitting $\xi$. These collected structures encompass nearly all commonly used elements across the periodic table. As shown in Fig. 3b, the elemental distribution within the dataset spans most of the periodic table, ensuring broad chemical diversity.

Having established a partitioned training approach for Hamiltonian components, it is important to note that accurate Hamiltonian reconstruction constitutes merely the first step toward achieving reliable band structure prediction. Directly incorporating band energy errors into the loss function can lead to optimization instabilities, as the gradient related to the eigenvalues of the Hamiltonian matrix with insufficient accuracy may diverge. We therefore adopt a two-stage training protocol (detailed in Methods) that first optimizes Hamiltonian fidelity before refining eigenvalue consistency. This sequential prioritization enhances numerical stability while preserving physical rigor.

An additional consideration for a universal model trained on diverse crystals is the handling of intrinsic variances in the zero points of DFT potential energy across different structures. Uncorrected, these system-dependent offsets introduce spurious translation terms into the Hamiltonian and eigenvalues, compromising predictions. To mitigate this, we systematically subtract the zero-point energy for each crystal using correction formulas provided in Methods. This renormalization step ensures that the training process focuses on physically meaningful energy differences rather than arbitrary absolute shifts, bolstering the model’s transferability across materials.

\begin{figure}[b]
\centering
\includegraphics[width=0.9\textwidth]{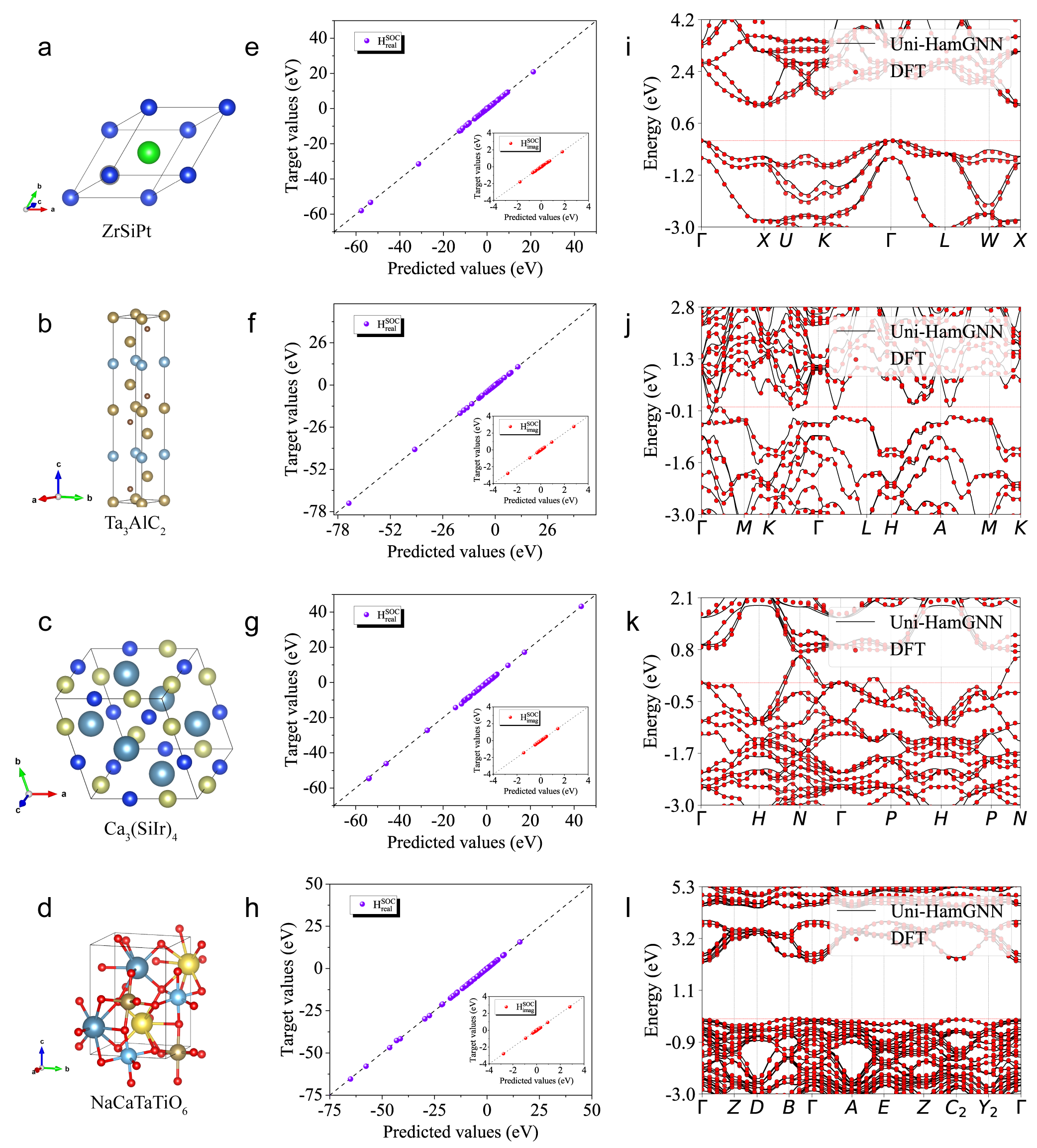}
\caption{\textbf{The prediction of Uni-HamGNN on several periodic solids that are not present in the training sets.} \textbf{a--d} Crystal structures of ZrSiPt, Ca$_3$(SiIr)$_4$, Ta$_3$AlC$_2$, and Na$_2$BiPCO$_7$. \textbf{e--h} Comparison of the Hamiltonian matrix elements predicted by Uni-HamGNN and those calculated by OpenMX for ZrSiPt, Ca$_3$(SiIr)$_4$, Ta$_3$AlC$_2$, and Na$_2$BiPCO$_7$. \textbf{i--l} Comparison of the energy bands predicted by Uni-HamGNN (\textit{solid line}) and those (\textit{dashed line}) calculated by OpenMX for ZrSiPt, Ca$_3$(SiIr)$_4$, Ta$_3$AlC$_2$, and Na$_2$BiPCO$_7$.}\label{fig4}
\end{figure}

\subsection{Performance evaluation on the test set}\label{subsec2}
To systematically assess the generalization capability of our trained universal SOC Hamiltonian model, we constructed an independent test set comprising 5,000 materials spanning diverse chemical compositions and crystal structures. For these systems, we calculated both the ground-truth SOC Hamiltonian matrices and their model-predicted counterparts. The model exhibited strong predictive accuracy, achieving a mean absolute error (MAE) of 3.58~$\mathrm{meV}$ for the real part of the whole SOC Hamiltonian ($H_{\mathrm{real}}^{\mathrm{SOC}}$) and an exceptionally low MAE of 0.0025~$\mathrm{meV}$ for the imaginary component ($H_{\mathrm{imag}}^{\mathrm{SOC}}$). These results confirm that our parameterization approach effectively quantifies SOC interactions with high fidelity. The parity plots in Figs. 3c and 3d further corroborate this conclusion, demonstrating excellent agreement between predicted and DFT-computed matrix elements across both real and imaginary components.

After confirming the model's overall predictive accuracy, we conducted detailed analyses on four representative heavy-element systems from the test set to further assess the model's reliability in complex scenarios: \text{ZrSiPt}, \text{Ca\textsubscript{3}(SiIr)\textsubscript{4}}, \text{Ta\textsubscript{3}AlC\textsubscript{2}}, and \text{Na\textsubscript{2}BiPCO\textsubscript{7}}. These materials were specifically chosen due to their pronounced SOC effects, which impose stringent demands on predictive accuracy. For the real components, the MAEs remained remarkably low at 1.29~$\mathrm{meV}$, 1.82~$\mathrm{meV}$, 1.51~$\mathrm{meV}$, and 1.22~$\mathrm{meV}$, respectively. The MAEs for the imaginary parts were found to be minuscule, at 0.0038~$\mathrm{meV}$, 0.0017~$\mathrm{meV}$, 0.0034~$\mathrm{meV}$, and 0.0022~$\mathrm{meV}$, respectively. This consistency across chemically distinct systems highlights the model’s capability to resolve subtle SOC-induced energy splits---a prerequisite for applications in spintronics and topological materials.

To validate the physical meaningfulness of these predictions, we compared the model's band structure calculations against full DFT results. As illustrated in Fig. 4, the universal model reproduces key SOC-induced features with exceptional accuracy, including band gap evolution, spin splitting patterns, and edge state modifications. Validations on more heavy-element systems are provided in the Supplementary Materials. The high accuracy demonstrated for Hamiltonian components, along with the excellent agreement with DFT-derived band structures, confirms the model's robustness and broad applicability in precisely describing SOC-related physical properties across a wide variety of complex material systems.

\subsection{High-Throughput Topological Insulator Screening}\label{subsec2}
Topological insulators (TIs) exhibit nontrivial topological properties primarily due to SOC effect. Crucially, SOC drives topological phase transitions through band inversion, governs spin-momentum locking in surface states, and underpins their distinctive quantum behavior\cite{bib4,bib7,bib27}. To distinguish TIs from trivial insulators, the $Z_2$ invariant---a topological index defined for non-magnetic systems with time-reversal symmetry---serves as a key discriminator. The $Z_2$ values of 0 and 1 correspond to trivial and nontrivial topological phases, respectively. Within the tight-binding framework, the $Z_2$ invariant is often evaluated across a series of 2D time-reversal invariant planes to characterize the band topology of three-dimensional materials. For a given 2D plane, the $Z_2$ invariant is expressed as\cite{bib28}:

\begin{figure}[t]
\centering
\includegraphics[width=0.9\textwidth]{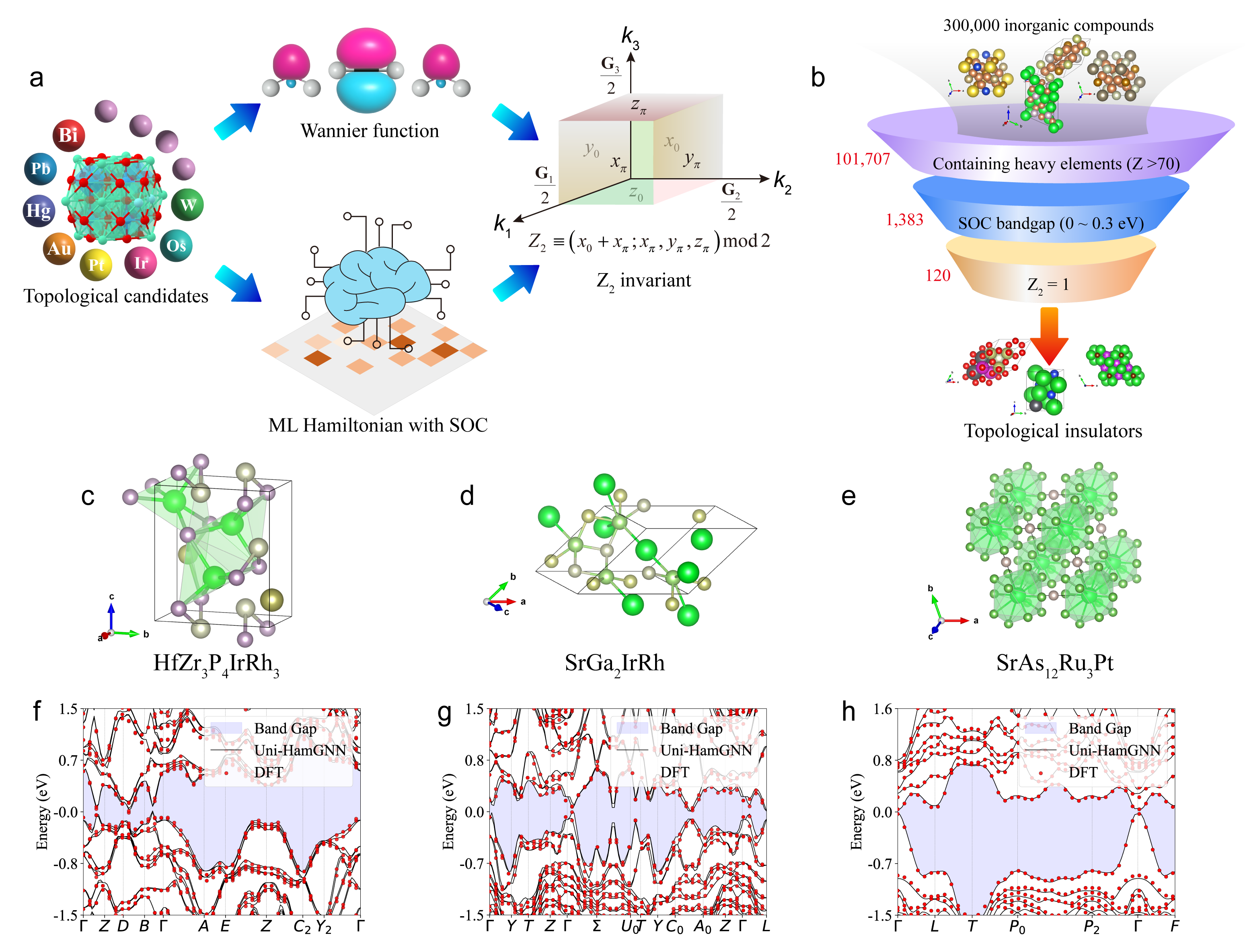}
\caption{\textbf{High-throughput topological insulator screening framework based on Uni-HamGNN.} \textbf{a} Schematic illustrating the calculation of $Z_2$ topological invariants using Wannier function-based methods versus the machine-learned Hamiltonian approach. \textbf{b} Workflow for high-throughput $Z_2$ topological insulator screening within the GNoME materials database. \textbf{c--e} Schematic crystal structures of HfZr$_3$P$_4$IrRh$_3$ (\textbf{c}), SrGa$_2$IrRh (\textbf{d}), and SrAs$_{12}$Ru$_3$Pt (\textbf{e}). \textbf{f--h} Comparison between Uni-HamGNN-predicted and DFT-calculated band structures with SOC for HfZr$_3$P$_4$IrRh$_3$ (\textbf{f}), SrGa$_2$IrRh (\textbf{g}), and SrAs$_{12}$Ru$_3$Pt (\textbf{h}).}\label{fig5}
\end{figure}

\begin{equation}
Z_2 = \frac{1}{2\pi} \sum_{n}^{\mathrm{occ.}} \left( \int_{\partial B} \boldsymbol{A}_n \cdot d\boldsymbol{k} - \int_{B} \Omega_n^{\alpha\beta} \, dk_\alpha dk_\beta \right) \pmod{2}
\tag{5}
\end{equation}

\noindent Here, $\boldsymbol{A}_n = -i \langle u_{n\boldsymbol{k}} | \frac{\partial}{\partial \boldsymbol{k}} | u_{n\boldsymbol{k}} \rangle$ represents the Berry connection, while $\boldsymbol{\Omega}_n = \nabla \times \boldsymbol{A}_n$ denotes the Berry curvature. $u_{n\boldsymbol{k}} = e^{-i\boldsymbol{k} \cdot \boldsymbol{r}} \psi_{n\boldsymbol{k}}$ is the lattice-periodic Bloch wavefunction. In three-dimensional systems, the Brillouin zone contains six time-reversal invariant planes ($k_i = 0$ and $k_i = \frac{\boldsymbol{G}_i}{2}$ for $i = 1,2,3$), yielding six $Z_2$ invariants $(x_0, x_\pi, y_0, y_\pi, z_0, z_\pi)$.

While the $Z_2$ framework provides a rigorous classification scheme, determining these invariants experimentally or computationally remains challenging. The conventional approach involves constructing tight-binding models from Wannier functions\cite{bib29,bib30}, which demands substantial computational resources and iterative parameter tuning (Fig. 5a). This method, though widely adopted, becomes impractical for large-scale material screening. A more efficient alternative lies in employing a universal SOC Hamiltonian model, which bypasses the need for expensive DFT calculations and Wannier basis construction. By directly providing the real-space SOC Hamiltonian matrix, this approach enables rapid computation of topological invariants such as the $Z_2$ index.

To leverage these computational advantages, we performed high-throughput screening of 10,170 heavy-element ($Z > 70$) materials from the GNoME\cite{bib31} dataset using Uni-HamGNN. Heavy elements were prioritized due to their pronounced SOC effects, which strongly influence topological phase transitions. Fig. 5b outlines our screening workflow: After identifying 1,383 insulating candidates with band gaps $\leq 0.3~\mathrm{eV}$---a range where SOC-induced band inversions are more likely to produce nontrivial states---we computed $Z_2$ invariants for these materials. This process revealed 120 topological insulators, demonstrating the method’s efficacy in discovering novel TI candidates.

Three representative systems from this set---\text{HfZr\textsubscript{3}P\textsubscript{4}IrRh\textsubscript{3}}, \text{SrGa\textsubscript{2}IrRh}, and \text{SrAs\textsubscript{12}Ru\textsubscript{3}Pt}---illustrate the predictive power of our approach (Figs. 5c--e). Their SOC band structures, calculated via both the universal model and DFT methods (Figs. 5f--h), show remarkable agreement, validating the model’s accuracy. The predicted $Z_2$ indices $(0,0,1,0,1,0)$, $(1,0,1,0,0,0)$, and $(1,0,1,0,1,0)$ were further confirmed through Wannier charge center evolution analysis using VASP (see Supplementary Materials). This consistency between model predictions and independent DFT-based verification not only underscores the reliability of our universal Hamiltonian approach but also establishes its potential as a powerful tool for accelerated TI discovery.

\begin{figure}[t]
\centering
\includegraphics[width=0.9\textwidth]{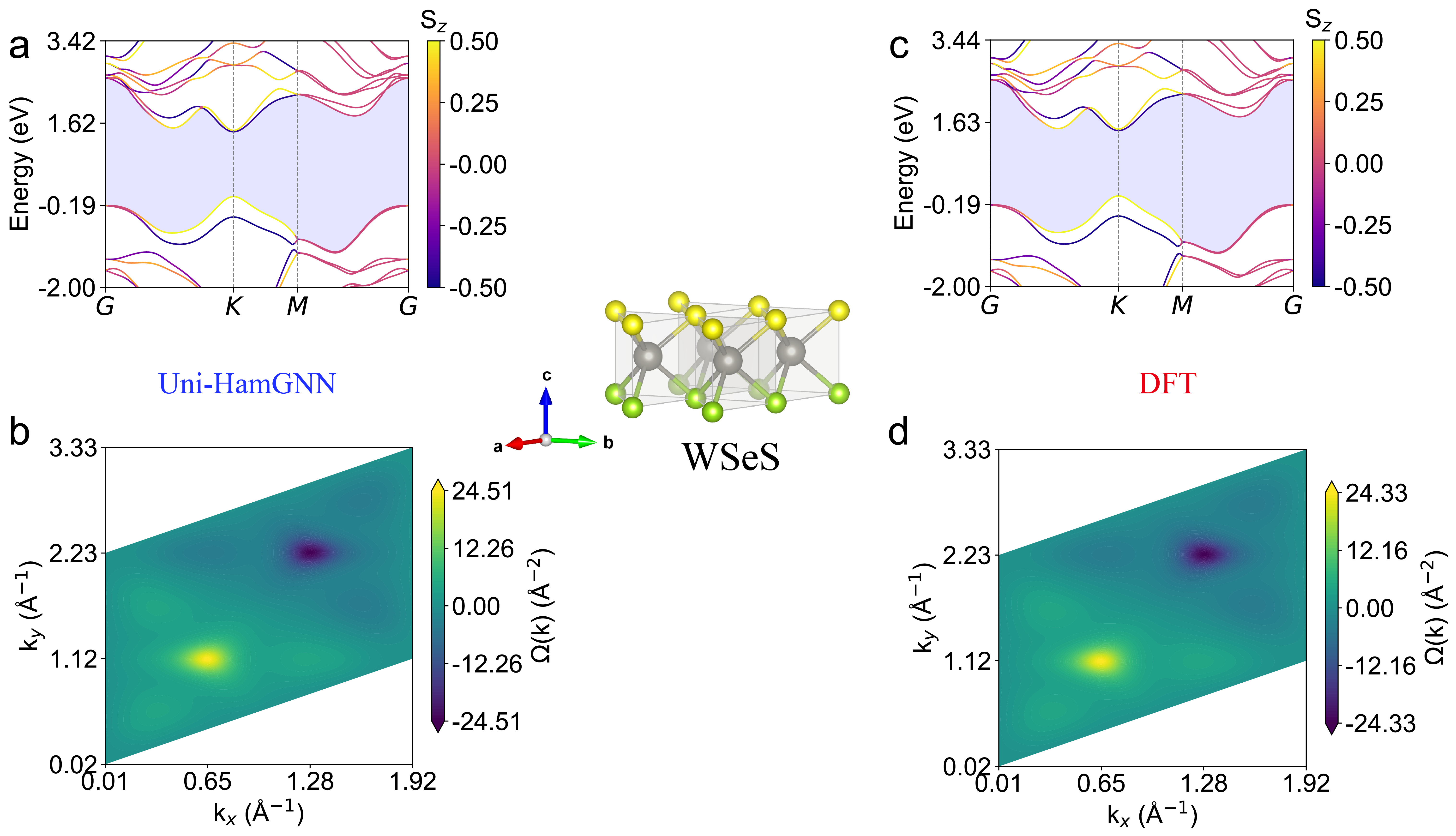}
\caption{\textbf{Comparison of SOC band structures and Berry curvature distributions for WSeS calculated by Uni-HamGNN \textit{vs} DFT.} \textbf{a} SOC band structure of WSeS computed via Uni-HamGNN. \textbf{b} Berry curvature distribution of WSeS computed via Uni-HamGNN. \textbf{c} SOC band structure of WSeS calculated by DFT. \textbf{d} Berry curvature distribution of WSeS calculated by DFT.}\label{fig6}
\end{figure}

\begin{figure}[t]
\centering
\includegraphics[width=0.9\textwidth]{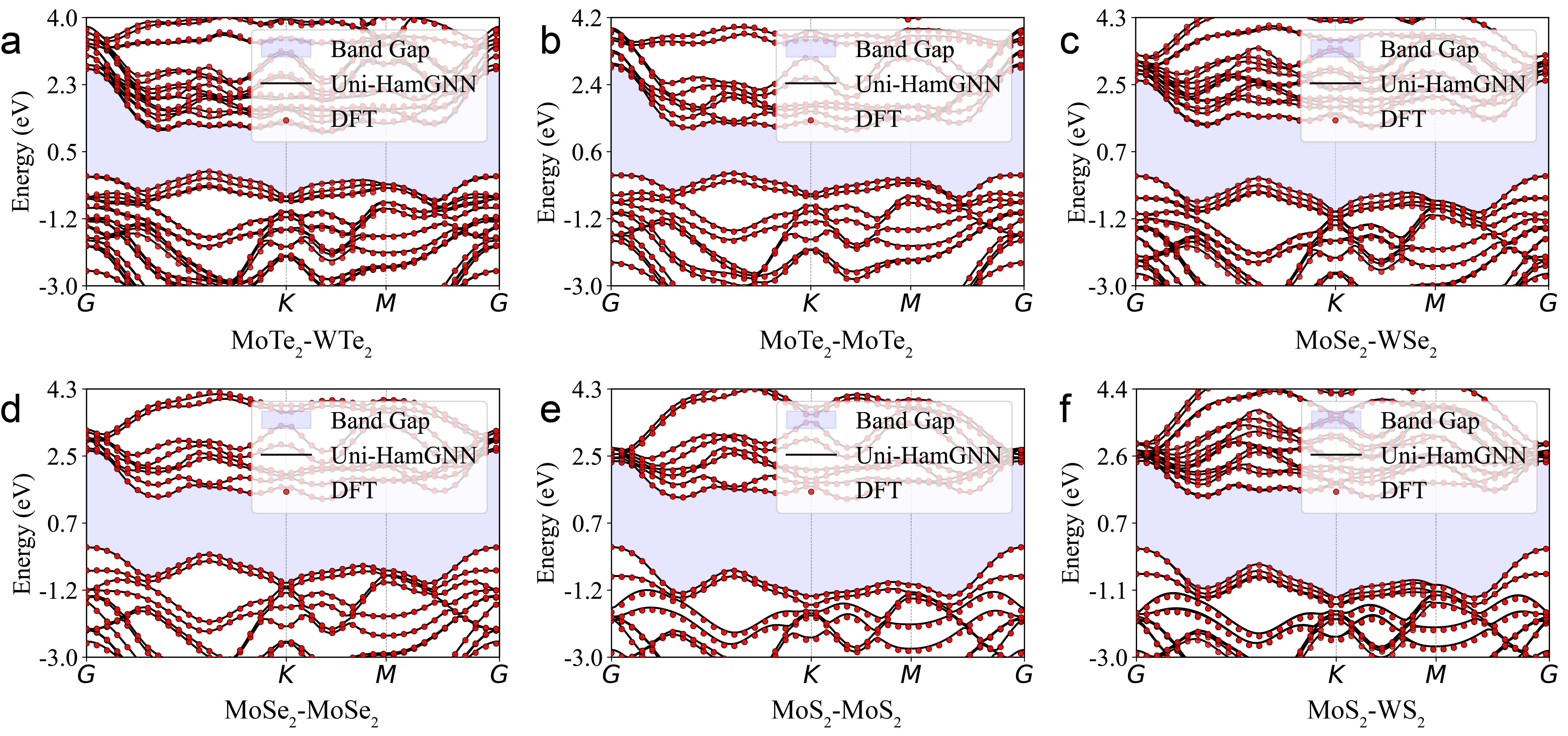}
\caption{\textbf{Comparison of spin-orbit coupling (SOC) band structures between Uni-HamGNN and DFT calculations for bilayer heterostructures of transition metal dichalcogenides MX$_2$ (M = Mo, W; X = S, Se, Te).}}\label{fig7}
\end{figure}

\begin{figure}[t]
\centering
\includegraphics[width=0.9\textwidth]{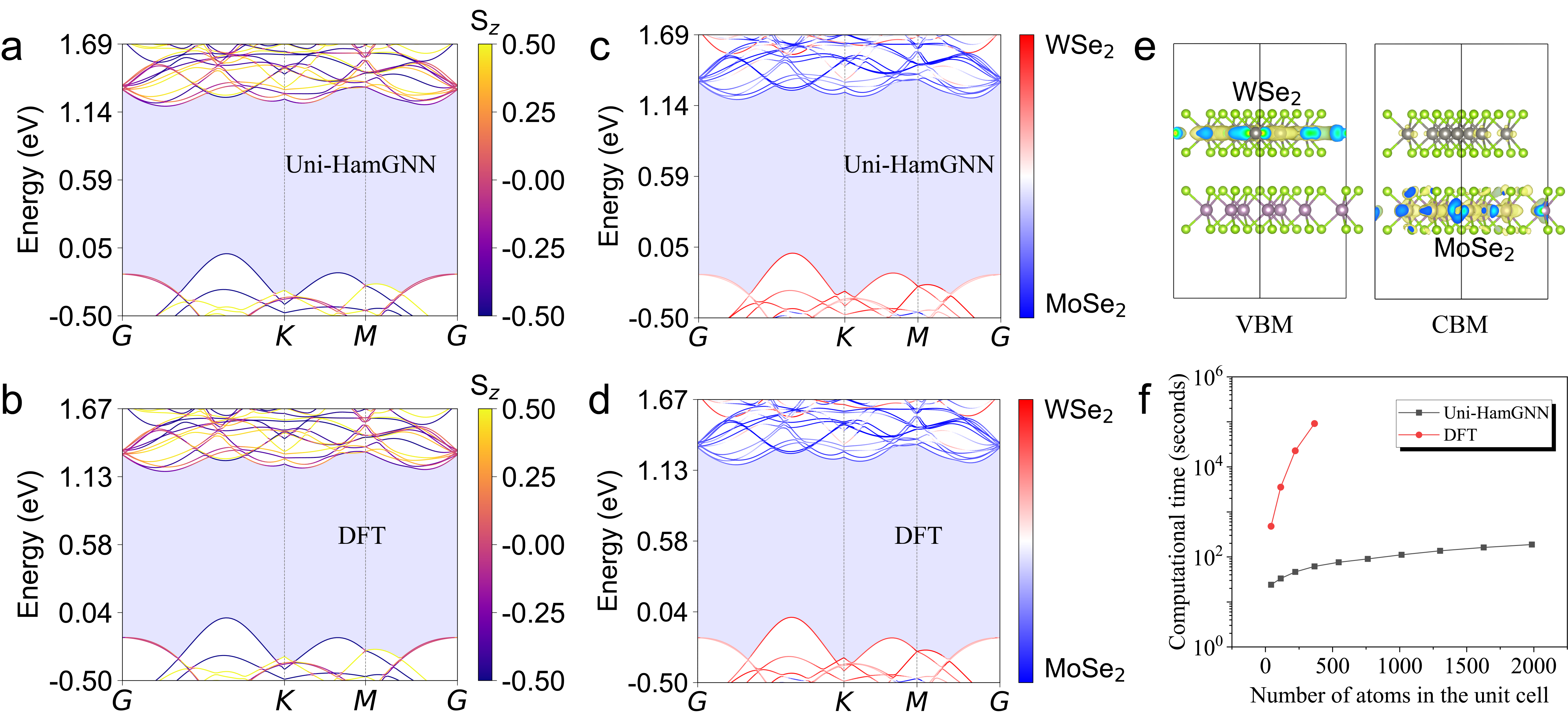}
\caption{\textbf{Comparison between Uni-HamGNN and DFT calculations for the MoSe$_2$-WSe$_2$ bilayer with a twist angle of $21.79^\circ$.} \textbf{a} Spin-projected band structures computed by Uni-HamGNN and \textbf{b} corresponding DFT results. \textbf{c} Orbital contribution distributions of MoSe$_2$-WSe$_2$ in the band structures predicted by Uni-HamGNN and \textbf{d} those calculated by DFT. \textbf{e} Charge density distributions at the valence band maximum (VBM) and conduction band minimum (CBM) predicted by Uni-HamGNN. \textbf{f} Computational time comparison between Uni-HamGNN and DFT for MoSe$_2$-WSe$_2$ bilayers with varying twist angles using 64 CPU cores.}\label{fig8}
\end{figure}

\subsection{2D Valleytronics and heterostructure}\label{subsec2}
While Uni-HamGNN is trained on three-dimensional (3D) bulk materials from the Materials Project database, its exceptional transferability enables the model to accurately calculate the SOC electronic structures of two-dimensional (2D) materials. This capability positions Uni-HamGNN as a highly efficient and reliable tool for exploring the electronic structures, topological properties, and other characteristics of 2D materials, including 2D valleytronics materials and bilayer heterostructures.

Valleytronics uses the ``valleys'' in the electronic band structure of solids as new degrees of freedom for information encoding, enabling lower energy consumption for data storage and transmission compared to traditional charge or spin-based devices\cite{bib8,bib9}. However, realizing practical valleytronic devices requires materials with stringent characteristics: multiple inequivalent valleys, direct bandgaps, and strong spin--orbit coupling to enable spin-valley locking. Addressing these challenges demands high-throughput screening tools to accelerate the discovery of viable candidates. In this work, we leverage Uni-HamGNN to predict SOC Hamiltonians for non-centrosymmetric 2D materials from the C2DB database\cite{bib32,bib33}, such as \text{WSeS}, \text{Ge\textsubscript{3}Bi\textsubscript{2}O\textsubscript{9}}, \text{InSeS\textsubscript{3}}, \text{TaAgP\textsubscript{2}Se\textsubscript{6}}, and \text{PbCl\textsubscript{2}}. From these predictions, we derived band structures and Berry curvature distributions across the first Brillouin zone. As exemplified by the Janus material \text{WSeS} in Fig. 6, Uni-HamGNN achieves excellent agreement with DFT in both SOC band structures (Figs. 6a and 6c) and Berry curvature profiles (Figs. 6b and 6d). The model’s spin-projection results align closely with DFT, confirming its accuracy in resolving spin-valley locking. Notably, the Berry curvature extrema at opposing valleys---critical for valley polarization---are precisely captured, underscoring Uni-HamGNN's ability to resolve subtle SOC-driven phenomena.

Expanding beyond monolayer systems, we investigated interfacial SOC effects in bilayer TMD heterostructures. Interlayer coupling in stacked TMDs can renormalize band edges and spin splittings---effects critical for tailoring quantum transport. As Fig. 7 illustrates, Uni-HamGNN faithfully reproduces DFT-derived band structures (solid vs. dashed lines) across all MX\textsubscript{2} (M = \text{Mo}, \text{W}; X = \text{S}, \text{Se}, \text{Te}) combinations, highlighting the model's capacity to simulate interfacial quantum phenomena with first-principles fidelity.

To further validate the robustness of Uni-HamGNN, we examined twisted bilayer systems, which not only exhibit intricate electronic structures but also impose significant computational challenges, especially for small-angle twisted configurations. As a representative case, we focused on twisted \text{MoSe\textsubscript{2}-WSe\textsubscript{2}} bilayers and calculated their band structures across varying twist angles using Uni-HamGNN. Figs. 8a and 8b present a direct comparison between the spin-projected band structures predicted by Uni-HamGNN and those obtained from DFT calculations, revealing strikingly consistent electronic features across both methodologies. Beyond band structure agreement, we analyzed the layer-specific orbital contributions to validate the model's physical interpretability. As illustrated in Figs. 8c and 8d, the orbital weights attributed to \text{MoSe\textsubscript{2}} and \text{WSe\textsubscript{2}} layers---derived from both Uni-HamGNN and DFT---show excellent quantitative alignment. Notably, this consistency extends to the spatial charge distribution characteristics, as illustrated in Fig. 8e. The holes at the valence band maximum (VBM) are primarily distributed in the \text{WSe\textsubscript{2}} layer, and the electrons at the conduction band minimum (CBM) are mainly located in the \text{MoSe\textsubscript{2}} layer.

Beyond accuracy, the computational efficiency of Uni-HamGNN represents a transformative advantage. Fig. 8f quantifies the dramatic reduction in computation time compared to DFT across different twist angles. Specifically, Uni-HamGNN achieves speedups of 2--3 orders of magnitude while maintaining high fidelity. Such efficiency gains establish the method as a practical and powerful tool for high-throughput exploration of quantum materials, where traditional DFT calculations are hindered by prohibitively high computational costs.

\section{Discussion}\label{sec1}
This study introduces a universal machine-learning framework (Uni-HamGNN) for predicting SOC Hamiltonians across the periodic table. By decomposing the SOC Hamiltonian into spin-independent and SOC correction terms, our approach preserves $\mathrm{SU}(2)$ symmetry while significantly reducing computational complexity. Building on this decomposition, we develop a delta-learning strategy that leverages abundant non-SOC Hamiltonian data alongside sparse SOC datasets, effectively lowering the training cost and difficulty of constructing a universal model. Rigorous validation against DFT calculations on a test set of 5,000 materials confirms Uni-HamGNN's high precision, robustness, and reliability. Furthermore, the framework's utility in high-throughput screening is demonstrated through its successful identification of topological insulators and its applications to valleytronic materials and TMD heterostructures. These results represent a meaningful leap forward in quantum material design, offering a computationally efficient alternative to conventional DFT-based approaches.

The enhanced transferability of Uni-HamGNN---compared to prior machine-learning methods restricted to specific material system---stems primarily from its delta-learning strategy and two-stage training protocol. The delta-learning approach addresses the inherent challenge of training a universal model caused by the orders-of-magnitude disparity between non-SOC terms and SOC corrections, thereby simplifying model complexity and enabling large-scale training across diverse materials. Importantly, even when the Hamiltonian itself achieves low prediction errors, the corresponding band structure may still exhibit significant deviations without proper regularization. This issue is resolved by our two-stage training method, which incorporates band structure errors in $k$-space as a regularization term during fine-tuning. This step not only mitigates overfitting but also steers the predicted Hamiltonian toward physically meaningful configurations, ensuring consistency between the Hamiltonian and its derived electronic properties.

While Uni-HamGNN demonstrates significant advantages, practical implementation considerations must be addressed. Universal models typically require large parameter counts to achieve high accuracy, posing challenges for deployment on low-memory GPUs. Although such models surpass dedicated single-system models in generalizability, their computational demands remain substantial. Recent advances in model distillation techniques offer a promising solution: by compressing the universal model via knowledge distillation, we can generate lightweight, efficient variants compatible with resource-constrained hardware. Beyond computational optimization, extending Uni-HamGNN to magnetic systems represents a critical frontier. The delta-learning framework and two-stage training methodology proposed here provide a natural pathway for this extension. Specifically, the magnetic Hamiltonian could be decomposed into non-magnetic and magnetic correction terms, analogous to our treatment of SOC effects. Integrating this decomposition with the current training strategy would enable a unified framework for predicting both SOC and magnetic Hamiltonians, significantly broadening the scope of quantum material simulations.
\section{Methods}\label{sec1}
\subsection{Network hyperparameters}\label{subsec2}
The equivariant node features are constructed through a combination of irreducible representations with specified rotational orders ($l$) and parity symmetries. Specifically, the feature composition follows: 64×0e + 64×0o + 32×1o + 16×1e + 12×2o + 25×2e + 18×3o + 9×3e + 4×4o + 9×4e + 4×5o + 4×5e + 2×6e. Here, ``64×0e'' denotes 64 feature channels, where each channel corresponds to an $O(3)$ irreducible representation with rotational order $l=0$ and odd parity. To encode atomic identity, each atom is represented as a one-hot vector of length 128 based on its atomic number. For spatial embedding, interatomic directions are projected onto a basis of real spherical harmonics with a maximum rotational order of $l_{\text{max}} = 5$.

The Uni-HamGNN architecture integrates $T=3$ orbital convolution layers, self-interaction layers, and pairwise interaction layers. The atomic neighborhood for each atom is defined by a cutoff radius aligned with the spatial extent of its atomic orbital basis. To ensure smooth radial representations, interatomic distances are expanded using 128 Bessel basis functions:
\begin{equation}
B\left(\left\|\mathbf{r}_{ij}\right\|\right)=\sqrt{\frac{2}{r_c}}\frac{\sin\left(\frac{n\pi\left\|\mathbf{r}_{ij}\right\|}{r_c}\right)}{\left\|\mathbf{r}_{ij}\right\|}f_{\mathrm{cutoff}}\left(\left\|\mathbf{r}_{ij}\right\|\right)
\tag{6}
\end{equation}
Here, $f_{\mathrm{cutoff}}$ is a cosine truncation function\cite{bib34} that enforces continuity for atoms near the cutoff boundary. To generate scalar weight coefficients $S_{c_o}^{(l_xl_fl_o)}\left(\left\|\mathbf{r}_{ij}\right\|\right)$ for tensor product operations, the radial basis $B\left(\left\|\mathbf{r}_{ij}\right\|\right)$ is processed through a two-layer multilayer perceptron (MLP) with 128 neurons per layer. 

\subsection{Optimization of Tensor Product Operations}\label{subsec2}
In the orbital convolution and pair interaction layers of the network, atomic and bond features are updated via tensor-product operations. These operations not only constitute the majority of the network parameters but also play a critical role in determining both computational efficiency and model expressiveness. Generally, tensor-product convolution can be represented by the following expression\cite{bib35,bib36}:

\begin{equation}
\begin{aligned}
M_{c_0l_om_o}^{ij} = \sum_{(l_xl_fl_o)}^{N_p} \sum_{c_x}^{C_x} \Biggl\{ & S_{c_xc_o}^{(l_xl_fl_o)}(\|\mathbf{r}_{ij}\|) \\
& \times \sum_{m_x=-l_x}^{l_x} \sum_{m_f=-l_f}^{l_f} \left( C_{m_xm_fm_o}^{l_xl_fl_o} \, m_{c_xl_xm_x} \, Y(\hat{\mathbf{r}}_{ij})_{l_fm_f} \right) \Biggr\}
\end{aligned}
\tag{7}
\end{equation}

\noindent In this formula, $m_{c_x l_x m_x}$ encodes either the atomic features $V_{c_x l_x m_x}^i \parallel V_{c_x l_x m_x}^j$ or the bond features $P_{c_x l_x m_x}^{ij}$. The notation $(l_x l_f l_o)$ enumerates all valid tensor-product paths $l_x \otimes l_f \rightarrow l_o$, where $N_p$ represents the total number of such paths. The radial function $S_{c_x c_o}^{(l_x l_f l_o)}\left(|\mathbf{r}_{ij}|\right) = \text{MLP}\left(B\left(|\mathbf{r}_{ij}|\right)\right)$ generates channel-dependent scalar weights through a multilayer perceptron (MLP), while $C_{m_x m_f m_o}^{l_x l_f l_o}$ denotes the Clebsch-Gordan coefficients governing rotational order coupling.

While the tensor product described in Eq. (7) is widely utilized in many equivariant networks, its computational demands become prohibitive for a universal Hamiltonian model. This complexity stems from the need for a universal model to incorporate a broad range of atomic feature channels (i.e., large $C_o$ and $C_x$) to improve expressiveness and generalization. Furthermore, in the context of a Hamiltonian matrix model, the rotational order $l$ can reach up to 5 or 6, leading to a substantial increase in the number of tensor-product paths ($N_p$). Consequently, the volume of learnable parameters in Eq. (7) expands rapidly, proportional to $N_p \times C_o \times C_x$, placing a heavy burden on GPU memory. To mitigate this computational challenge, we have refined Eq. (7) by adopting an optimized tensor-product format in HamGNN-V2:

\begin{equation}
\begin{aligned}
M_{c_0 l_o m_o}^{ij} = & \sum_{(l_x l_f l_o)}^{N_p} S_{c_o}^{(l_x l_f l_o)}(|\mathbf{r}_{ij}|) 
\sum_{c_x}^{C_x} \sum_{m_x=-l_x}^{l_x} \sum_{m_f=-l_f}^{l_f} \\
& \Biggl( W_{c_o c_x} \biggl( 
C_{m_x m_f m_o}^{l_x l_f l_o} \, m_{c_x l_x m_x} 
\, Y({\hat{\mathbf{r}}}_{ij})_{l_f m_f} 
\biggr) \Biggr)
\end{aligned}
\tag{8}
\end{equation}

In Eq. (8), the tensor-product of $m_{c_x l_x m_x}$ and $Y\left({\hat{\mathbf{r}}}_{ij}\right)_{l_f m_f}$ is first linearly transformed via a weight matrix $W_{c_o c_x}$, which maps the input channel dimension $C_x$ to the output channel dimension $C_o$. Subsequently, an equivariant scaling is applied to each tensor-product path using a scalar coefficient $S_{c_o}^{(l_x l_f l_o)}\left(|\mathbf{r}_{ij}|\right)$. This modification substantially decreases the number of learnable parameters. Specifically, the parameter count for $W_{c_o c_x}$ is $C_o \times C_x$, and the parameter count for $S_{c_o}^{(l_x l_f l_o)}\left(|\mathbf{r}_{ij}|\right)$ is $N_p \times C_o$. Compared to Eq. (7), the total number of parameters required by Eq. (8) is only $\frac{1}{N_p} + \frac{1}{C_x}$ of the former, thus significantly reducing computational resource demands while preserving the model's expressiveness.

\subsection{Zero-point renormalization}\label{subsec2}
In different unit cells, the zero point of potential energy may vary, leading to a translation term between the model-predicted Hamiltonian $\widetilde{H}$ and the DFT-computed Hamiltonian $H$: $\widetilde{H} = H + \theta \cdot S$, where $\theta$ is a constant and $S$ is the overlap matrix. This translation term can interfere with the training process and must therefore be eliminated. To mitigate the impact of the zero-point drift on the training of the real-space Hamiltonian, we applied the following correction to the model-predicted Hamiltonian $\widetilde{H}$:

\begin{equation}
{\widetilde{H}}^\prime = \widetilde{H} - \frac{\sum\limits_{ij} \left( {\widetilde{H}}_{ij} - H_{ij} \right)}{\sum\limits_{ij} S_{ij}} S_{ij}
\tag{9}
\end{equation}

Additionally, the zero-point drift in potential energy can introduce a constant shift between the model-predicted band eigenvalues ${\widetilde{E}}_{nk}$ and the DFT-computed band eigenvalues $E_{nk}$. To address this issue, we corrected the predicted band eigenvalues using the following formula:

\begin{equation}
{\widetilde{E}}_{nk}^\prime = {\widetilde{E}}_{nk} - \frac{1}{N_k N_b} \left( {\widetilde{E}}_{nk} - E_{nk} \right)
\tag{10}
\end{equation}

\subsection{Two-stage training protocol}\label{subsec2}
Directly incorporating band energy errors into the loss function can lead to optimization instabilities, as gradients derived from the eigenvalues of an inaccurately predicted Hamiltonian matrix may diverge. To circumvent this challenge while maintaining physical consistency, we implement a two-stage training protocol that systematically prioritizes Hamiltonian fidelity before refining eigenvalue agreement. This hierarchical approach ensures stable gradient propagation while aligning predictions with both real-space and reciprocal-space observables.

In the first training stage, the loss function was defined as the error between the model-predicted real-space Hamiltonian ${\widetilde{H}}^\prime$ and the DFT-computed Hamiltonian $H$, aiming to make the model closely approximate the true physical Hamiltonian in real space. Specifically, the loss function for the first training stage is defined as:

\begin{equation}
L_{\text{stage1}} = \lVert H' - H \rVert
\tag{11}
\end{equation}

\noindent where $\lVert \cdot \rVert$ represents the mean absolute error of each matrix element. In this stage, we trained the model until convergence to ensure that ${\widetilde{H}}^\prime$ closely matches $H$ in real space. This stage can be viewed as an initial calibration of the model, laying the foundation for subsequent fine-tuning.

Once the first training stage was completed, we loaded the parameters of the converged model and proceeded to the second training stage. In this stage, the reciprocal-space band structure constraints are introduced to refine eigenvalue accuracy. Specifically, the model-predicted Hamiltonian $\widetilde{H}$ was diagonalized to obtain the predicted band structure in $k$-space, ${\widetilde{E}}_{nk}^\prime$, which was then compared with the DFT-computed band structure $E_{nk}$. The loss function for this step is defined as:

\begin{equation}
L_{\text{stage2}} = \lVert H' - H \rVert + \frac{\lambda }{{{N_b} \times {N_k}}} \sum_{k=1}^{N_k} \sum_{n=1}^{N_b} \lvert E_{nk}' - E_{nk} \rvert
\tag{12}
\end{equation}

\noindent where $\lambda$ is the weight of the band structure error in the total loss, $N_k$ is the number of $k$-points randomly sampled in reciprocal space, and $N_b$ is the number of selected bands, typically focusing on those near the Fermi level. Given that the SOC terms ($\xi \widehat{\vec{L}} \cdot \widehat{\vec{\sigma}}$) act as a perturbation to the non-SOC component $H_0$ and exhibit negligible prediction errors after the first training stage, our second stage training focuses exclusively on the eigenvalues of $H_0$ without modifying SOC parameters.

Both training stages employ the AdamW\cite{bib37,bib38} optimizer with stage-dependent hyperparameters. Stage 1 utilizes an initial learning rate of $10^{-2}$ to efficiently navigate the parameter space, while Stage 2 adopts a reduced rate of $10^{-4}$ combined with $\lambda = 10^{-2}$ for precise eigenvalue tuning. To prevent overfitting, we implement an adaptive early stopping protocol: If validation accuracy plateaus for 5 consecutive epochs ($N_{\text{patience}} = 5$), the learning rate halves. Training terminates if no improvement persists for 30 epochs ($N_{\text{stop}} = 30$) or when the learning rate falls below $10^{-5}$, ensuring efficient resource utilization while maintaining model generalizability.

\subsection{The calculation of Z2 invariant}\label{subsec2}
For calculating the $Z_2$ topological number on each two-dimensional plane, the overlap matrix $U$ proposed by Fukui-Hatsugai-Suzuki can be employed, defined as\cite{bib28}:

\begin{equation}
U_{\Delta \mathbf{k}} = \det \left\langle u_n(\mathbf{k}) \middle| u_m(\mathbf{k} + \Delta \mathbf{k}) \right\rangle
\tag{13}
\end{equation}

\noindent Using the overlap matrix, Berry connection and Berry curvature can be computed on each plaquette:

\begin{equation}
A_{ab} = \operatorname{Im} \ln U_{ab}
\tag{14}
\end{equation}

\begin{equation}
\Omega(\mathbf{k}) = \operatorname{Im} \ln \left( U_{12} U_{23} U_{34} U_{41} \right)
\tag{15}
\end{equation}

\noindent The integer resonance $n \, (=0, \pm 1)$ on each plaquette can be calculated by:

\begin{equation}
n(\mathbf{k}) = \frac{1}{2\pi} \left( A_{12} + A_{23} + A_{34} + A_{41} - \Omega(\mathbf{k}) \right)
\tag{16}
\end{equation}

\noindent By summing $n$ across half of the Brillouin zone and taking the result modulo 2, the $Z_2$ invariant is obtained:

\begin{equation}
Z_2 = \frac{1}{2\pi} \sum_{\mathbf{k}}^{\text{Half BZ}} n(\mathbf{k}) \pmod{2}
\tag{17}
\end{equation}

\subsection{DFT calculation details}\label{subsec2}
DFT calculations were performed on Materials Project structures to generate real-space Hamiltonian matrices using OpenMX\cite{bib39}---a computational package designed for nanoscale material simulations utilizing norm-conserving pseudopotentials and pseudo-atomic localized basis orbitals. The training set Hamiltonians were computed with the following parameters: a $6 \times 6 \times 6$ Monkhorst-Pack $k$-point grid for Brillouin zone sampling, a SCF convergence threshold of $1.0 \times 10^{-8}~\mathrm{Hartree}$, and an energy cutoff of $200~\mathrm{Ry}$ for real-space discretization.

\section{Code and data availability}\label{sec1}
The HamGNN-V2 code is publicly accessible on GitHub at https://github.com/QuantumLab-ZY/HamGNN.  The network weights of Uni-HamGNN will be made publicly available once the paper is officially accepted.

\bibliography{sn-bibliography}


\begin{thebibliography}{41}
\ifx \bisbn   \undefined \def \bisbn  #1{ISBN #1}\fi
\ifx \binits  \undefined \def \binits#1{#1}\fi
\ifx \bauthor  \undefined \def \bauthor#1{#1}\fi
\ifx \batitle  \undefined \def \batitle#1{#1}\fi
\ifx \bjtitle  \undefined \def \bjtitle#1{#1}\fi
\ifx \bvolume  \undefined \def \bvolume#1{\textbf{#1}}\fi
\ifx \byear  \undefined \def \byear#1{#1}\fi
\ifx \bissue  \undefined \def \bissue#1{#1}\fi
\ifx \bfpage  \undefined \def \bfpage#1{#1}\fi
\ifx \blpage  \undefined \def \blpage #1{#1}\fi
\ifx \burl  \undefined \def \burl#1{\textsf{#1}}\fi
\ifx \doiurl  \undefined \def \doiurl#1{\url{https://doi.org/#1}}\fi
\ifx \betal  \undefined \def \betal{\textit{et al.}}\fi
\ifx \binstitute  \undefined \def \binstitute#1{#1}\fi
\ifx \binstitutionaled  \undefined \def \binstitutionaled#1{#1}\fi
\ifx \bctitle  \undefined \def \bctitle#1{#1}\fi
\ifx \beditor  \undefined \def \beditor#1{#1}\fi
\ifx \bpublisher  \undefined \def \bpublisher#1{#1}\fi
\ifx \bbtitle  \undefined \def \bbtitle#1{#1}\fi
\ifx \bedition  \undefined \def \bedition#1{#1}\fi
\ifx \bseriesno  \undefined \def \bseriesno#1{#1}\fi
\ifx \blocation  \undefined \def \blocation#1{#1}\fi
\ifx \bsertitle  \undefined \def \bsertitle#1{#1}\fi
\ifx \bsnm \undefined \def \bsnm#1{#1}\fi
\ifx \bsuffix \undefined \def \bsuffix#1{#1}\fi
\ifx \bparticle \undefined \def \bparticle#1{#1}\fi
\ifx \barticle \undefined \def \barticle#1{#1}\fi
\bibcommenthead
\ifx \bconfdate \undefined \def \bconfdate #1{#1}\fi
\ifx \botherref \undefined \def \botherref #1{#1}\fi
\ifx \url \undefined \def \url#1{\textsf{#1}}\fi
\ifx \bchapter \undefined \def \bchapter#1{#1}\fi
\ifx \bbook \undefined \def \bbook#1{#1}\fi
\ifx \bcomment \undefined \def \bcomment#1{#1}\fi
\ifx \oauthor \undefined \def \oauthor#1{#1}\fi
\ifx \citeauthoryear \undefined \def \citeauthoryear#1{#1}\fi
\ifx \endbibitem  \undefined \def \endbibitem {}\fi
\ifx \bconflocation  \undefined \def \bconflocation#1{#1}\fi
\ifx \arxivurl  \undefined \def \arxivurl#1{\textsf{#1}}\fi
\csname PreBibitemsHook\endcsname

\bibitem[\protect\citeauthoryear{Zutic et~al.}{2004}]{bib1}
\begin{barticle}
\bauthor{\bsnm{Zutic}, \binits{I.}},
\bauthor{\bsnm{Fabian}, \binits{J.}},
\bauthor{\bsnm{Das~Sarma}, \binits{S.}}:
\batitle{Spintronics: Fundamentals and applications}.
\bjtitle{Rev. Mod. Phys.}
\bvolume{76}(\bissue{2}),
\bfpage{323}--\blpage{410}
(\byear{2004})
\doiurl{10.1103/revmodphys.76.323}
\end{barticle}
\endbibitem

\bibitem[\protect\citeauthoryear{Dieny et~al.}{2020}]{bib2}
\begin{barticle}
\bauthor{\bsnm{Dieny}, \binits{B.}},
\bauthor{\bsnm{Prejbeanu}, \binits{I.L.}},
\bauthor{\bsnm{Garello}, \binits{K.}},
\bauthor{\bsnm{Gambardella}, \binits{P.}},
\bauthor{\bsnm{Freitas}, \binits{P.}},
\bauthor{\bsnm{Lehndorff}, \binits{R.}},
\bauthor{\bsnm{Raberg}, \binits{W.}},
\bauthor{\bsnm{Ebels}, \binits{U.}},
\bauthor{\bsnm{Demokritov}, \binits{S.O.}},
\bauthor{\bsnm{Akerman}, \binits{J.}},
\bauthor{\bsnm{Deac}, \binits{A.}},
\bauthor{\bsnm{Pirro}, \binits{P.}},
\bauthor{\bsnm{Adelmann}, \binits{C.}},
\bauthor{\bsnm{Anane}, \binits{A.}},
\bauthor{\bsnm{Chumak}, \binits{A.V.}},
\bauthor{\bsnm{Hirohata}, \binits{A.}},
\bauthor{\bsnm{Mangin}, \binits{S.}},
\bauthor{\bsnm{Valenzuela}, \binits{S.O.}},
\bauthor{\bsnm{Onbaşlı}, \binits{M.C.}},
\bauthor{\bsnm{d'Aquino}, \binits{M.}},
\bauthor{\bsnm{Prenat}, \binits{G.}},
\bauthor{\bsnm{Finocchio}, \binits{G.}},
\bauthor{\bsnm{Lopez-Diaz}, \binits{L.}},
\bauthor{\bsnm{Chantrell}, \binits{R.}},
\bauthor{\bsnm{Chubykalo-Fesenko}, \binits{O.}},
\bauthor{\bsnm{Bortolotti}, \binits{P.}}:
\batitle{Opportunities and challenges for spintronics in the microelectronics industry}.
\bjtitle{Nat. Electron.}
\bvolume{3}(\bissue{8}),
\bfpage{446}--\blpage{459}
(\byear{2020})
\doiurl{10.1038/s41928-020-0461-5}
\end{barticle}
\endbibitem

\bibitem[\protect\citeauthoryear{Sinova et~al.}{2015}]{bib3}
\begin{barticle}
\bauthor{\bsnm{Sinova}, \binits{J.}},
\bauthor{\bsnm{Valenzuela}, \binits{S.O.}},
\bauthor{\bsnm{Wunderlich}, \binits{J.}},
\bauthor{\bsnm{Back}, \binits{C.â.}},
\bauthor{\bsnm{Jungwirth}, \binits{T.}}:
\batitle{Spin hall effects}.
\bjtitle{Rev. Mod. Phys.}
\bvolume{87}(\bissue{4}),
\bfpage{1213}--\blpage{1260}
(\byear{2015})
\doiurl{10.1103/RevModPhys.87.1213}
\end{barticle}
\endbibitem

\bibitem[\protect\citeauthoryear{Qi and Zhang}{2011}]{bib4}
\begin{barticle}
\bauthor{\bsnm{Qi}, \binits{X.-L.}},
\bauthor{\bsnm{Zhang}, \binits{S.-C.}}:
\batitle{Topological insulators and superconductors}.
\bjtitle{Rev. Mod. Phys}
\bvolume{83}(\bissue{4}),
\bfpage{1057}--\blpage{1110}
(\byear{2011})
\doiurl{10.1103/RevModPhys.83.1057}
\end{barticle}
\endbibitem

\bibitem[\protect\citeauthoryear{Narang et~al.}{2020}]{bib5}
\begin{barticle}
\bauthor{\bsnm{Narang}, \binits{P.}},
\bauthor{\bsnm{Garcia}, \binits{C.A.C.}},
\bauthor{\bsnm{Felser}, \binits{C.}}:
\batitle{The topology of electronic band structures}.
\bjtitle{Nat. Mater.}
\bvolume{20}(\bissue{3}),
\bfpage{293}--\blpage{300}
(\byear{2020})
\doiurl{10.1038/s41563-020-00820-4}
\end{barticle}
\endbibitem

\bibitem[\protect\citeauthoryear{Xiao and Yan}{2021}]{bib6}
\begin{barticle}
\bauthor{\bsnm{Xiao}, \binits{J.}},
\bauthor{\bsnm{Yan}, \binits{B.}}:
\batitle{First-principles calculations for topological quantum materials}.
\bjtitle{Nat. Rev. Phys.}
\bvolume{3}(\bissue{4}),
\bfpage{283}--\blpage{297}
(\byear{2021})
\doiurl{10.1038/s42254-021-00292-8}
\end{barticle}
\endbibitem

\bibitem[\protect\citeauthoryear{Wang et~al.}{2020}]{bib7}
\begin{barticle}
\bauthor{\bsnm{Wang}, \binits{K.L.}},
\bauthor{\bsnm{Wu}, \binits{Y.}},
\bauthor{\bsnm{Eckberg}, \binits{C.}},
\bauthor{\bsnm{Yin}, \binits{G.}},
\bauthor{\bsnm{Pan}, \binits{Q.}}:
\batitle{Topological quantum materials}.
\bjtitle{MRS Bull.}
\bvolume{45}(\bissue{5}),
\bfpage{373}--\blpage{379}
(\byear{2020})
\doiurl{10.1557/mrs.2020.122}
\end{barticle}
\endbibitem

\bibitem[\protect\citeauthoryear{Schaibley et~al.}{2016}]{bib8}
\begin{barticle}
\bauthor{\bsnm{Schaibley}, \binits{J.R.}},
\bauthor{\bsnm{Yu}, \binits{H.}},
\bauthor{\bsnm{Clark}, \binits{G.}},
\bauthor{\bsnm{Rivera}, \binits{P.}},
\bauthor{\bsnm{Ross}, \binits{J.S.}},
\bauthor{\bsnm{Seyler}, \binits{K.L.}},
\bauthor{\bsnm{Yao}, \binits{W.}},
\bauthor{\bsnm{Xu}, \binits{X.}}:
\batitle{Valleytronics in 2d materials}.
\bjtitle{Nat. Rev. Mater.}
\bvolume{1}(\bissue{11}),
\bfpage{16055}
(\byear{2016})
\doiurl{10.1038/natrevmats.2016.55}
\end{barticle}
\endbibitem

\bibitem[\protect\citeauthoryear{Li et~al.}{2020}]{bib9}
\begin{barticle}
\bauthor{\bsnm{Li}, \binits{L.}},
\bauthor{\bsnm{Shao}, \binits{L.}},
\bauthor{\bsnm{Liu}, \binits{X.}},
\bauthor{\bsnm{Gao}, \binits{A.}},
\bauthor{\bsnm{Wang}, \binits{H.}},
\bauthor{\bsnm{Zheng}, \binits{B.}},
\bauthor{\bsnm{Hou}, \binits{G.}},
\bauthor{\bsnm{Shehzad}, \binits{K.}},
\bauthor{\bsnm{Yu}, \binits{L.}},
\bauthor{\bsnm{Miao}, \binits{F.}},
\bauthor{\bsnm{Shi}, \binits{Y.}},
\bauthor{\bsnm{Xu}, \binits{Y.}},
\bauthor{\bsnm{Wang}, \binits{X.}}:
\batitle{Room-temperature valleytronic transistor}.
\bjtitle{Nat. Nanotechnol.}
\bvolume{15}(\bissue{9}),
\bfpage{743}--\blpage{749}
(\byear{2020})
\doiurl{10.1038/s41565-020-0727-0}
\end{barticle}
\endbibitem

\bibitem[\protect\citeauthoryear{Jiang et~al.}{2019}]{bib10}
\begin{barticle}
\bauthor{\bsnm{Jiang}, \binits{S.}},
\bauthor{\bsnm{Li}, \binits{L.}},
\bauthor{\bsnm{Wang}, \binits{Z.}},
\bauthor{\bsnm{Shan}, \binits{J.}},
\bauthor{\bsnm{Mak}, \binits{K.F.}}:
\batitle{Spin tunnel field-effect transistors based on two-dimensional van der waals heterostructures}.
\bjtitle{Nat. Electron.}
\bvolume{2}(\bissue{4}),
\bfpage{159}--\blpage{163}
(\byear{2019})
\doiurl{10.1038/s41928-019-0232-3}
\end{barticle}
\endbibitem

\bibitem[\protect\citeauthoryear{Chuang et~al.}{2014}]{bib11}
\begin{barticle}
\bauthor{\bsnm{Chuang}, \binits{P.}},
\bauthor{\bsnm{Ho}, \binits{S.-C.}},
\bauthor{\bsnm{Smith}, \binits{L.W.}},
\bauthor{\bsnm{Sfigakis}, \binits{F.}},
\bauthor{\bsnm{Pepper}, \binits{M.}},
\bauthor{\bsnm{Chen}, \binits{C.-H.}},
\bauthor{\bsnm{Fan}, \binits{J.-C.}},
\bauthor{\bsnm{Griffiths}, \binits{J.P.}},
\bauthor{\bsnm{Farrer}, \binits{I.}},
\bauthor{\bsnm{Beere}, \binits{H.E.}},
\bauthor{\bsnm{Jones}, \binits{G.A.C.}},
\bauthor{\bsnm{Ritchie}, \binits{D.A.}},
\bauthor{\bsnm{Chen}, \binits{T.-M.}}:
\batitle{All-electric all-semiconductor spin field-effect transistors}.
\bjtitle{Nature Nanotech.}
\bvolume{10}(\bissue{1}),
\bfpage{35}--\blpage{39}
(\byear{2014})
\doiurl{10.1038/nnano.2014.296}
\end{barticle}
\endbibitem

\bibitem[\protect\citeauthoryear{Hegde and Bowen}{2017}]{bib12}
\begin{barticle}
\bauthor{\bsnm{Hegde}, \binits{G.}},
\bauthor{\bsnm{Bowen}, \binits{R.C.}}:
\batitle{Machine-learned approximations to density functional theory hamiltonians}.
\bjtitle{Sci. Rep.}
\bvolume{7},
\bfpage{42669}
(\byear{2017})
\doiurl{10.1038/srep42669}
\end{barticle}
\endbibitem

\bibitem[\protect\citeauthoryear{Schutt et~al.}{2019}]{bib13}
\begin{barticle}
\bauthor{\bsnm{Schutt}, \binits{K.T.}},
\bauthor{\bsnm{Gastegger}, \binits{M.}},
\bauthor{\bsnm{Tkatchenko}, \binits{A.}},
\bauthor{\bsnm{Muller}, \binits{K.R.}},
\bauthor{\bsnm{Maurer}, \binits{R.J.}}:
\batitle{Unifying machine learning and quantum chemistry with a deep neural network for molecular wavefunctions}.
\bjtitle{Nat. Commun.}
\bvolume{10},
\bfpage{5024}
(\byear{2019})
\doiurl{10.1038/s41467-019-12875-2}
\end{barticle}
\endbibitem

\bibitem[\protect\citeauthoryear{Unke et~al.}{2021}]{bib14}
\begin{botherref}
\oauthor{\bsnm{Unke}, \binits{O.T.}},
\oauthor{\bsnm{Bogojeski}, \binits{M.}},
\oauthor{\bsnm{Gastegger}, \binits{M.}},
\oauthor{\bsnm{Geiger}, \binits{M.}},
\oauthor{\bsnm{Smidt}, \binits{T.}},
\oauthor{\bsnm{Müller}, \binits{K.-R.}}:
SE(3)-equivariant prediction of molecular wavefunctions and electronic densities
(2021).
\url{https://ui.adsabs.harvard.edu/abs/2021arXiv210602347U}
\end{botherref}
\endbibitem

\bibitem[\protect\citeauthoryear{Zhang et~al.}{2022}]{bib15}
\begin{barticle}
\bauthor{\bsnm{Zhang}, \binits{L.}},
\bauthor{\bsnm{Onat}, \binits{B.}},
\bauthor{\bsnm{Dusson}, \binits{G.}},
\bauthor{\bsnm{McSloy}, \binits{A.}},
\bauthor{\bsnm{Anand}, \binits{G.}},
\bauthor{\bsnm{Maurer}, \binits{R.J.}},
\bauthor{\bsnm{Ortner}, \binits{C.}},
\bauthor{\bsnm{Kermode}, \binits{J.R.}}:
\batitle{Equivariant analytical mapping of first principles hamiltonians to accurate and transferable materials models}.
\bjtitle{npj Comput. Mater.}
\bvolume{8}(\bissue{1}),
\bfpage{158}
(\byear{2022})
\doiurl{10.1038/s41524-022-00843-2}
\end{barticle}
\endbibitem

\bibitem[\protect\citeauthoryear{Gong et~al.}{2023}]{bib16}
\begin{barticle}
\bauthor{\bsnm{Gong}, \binits{X.}},
\bauthor{\bsnm{Li}, \binits{H.}},
\bauthor{\bsnm{Zou}, \binits{N.}},
\bauthor{\bsnm{Xu}, \binits{R.}},
\bauthor{\bsnm{Duan}, \binits{W.}},
\bauthor{\bsnm{Xu}, \binits{Y.}}:
\batitle{General framework for e(3)-equivariant neural network representation of density functional theory hamiltonian}.
\bjtitle{Nat. Commun.}
\bvolume{14}(\bissue{1}),
\bfpage{2848}
(\byear{2023})
\doiurl{10.1038/s41467-023-38468-8}
\end{barticle}
\endbibitem

\bibitem[\protect\citeauthoryear{Zhong et~al.}{2023}]{bib17}
\begin{barticle}
\bauthor{\bsnm{Zhong}, \binits{Y.}},
\bauthor{\bsnm{Yu}, \binits{H.}},
\bauthor{\bsnm{Su}, \binits{M.}},
\bauthor{\bsnm{Gong}, \binits{X.}},
\bauthor{\bsnm{Xiang}, \binits{H.}}:
\batitle{Transferable equivariant graph neural networks for the hamiltonians of molecules and solids}.
\bjtitle{npj Comput. Mater.}
\bvolume{9}(\bissue{1}),
\bfpage{182}
(\byear{2023})
\doiurl{10.1038/s41524-023-01130-4}
\end{barticle}
\endbibitem

\bibitem[\protect\citeauthoryear{Gu et~al.}{2024}]{bib18}
\begin{barticle}
\bauthor{\bsnm{Gu}, \binits{Q.}},
\bauthor{\bsnm{Zhouyin}, \binits{Z.}},
\bauthor{\bsnm{Pandey}, \binits{S.K.}},
\bauthor{\bsnm{Zhang}, \binits{P.}},
\bauthor{\bsnm{Zhang}, \binits{L.}},
\bauthor{\bsnm{E}, \binits{W.}}:
\batitle{Deep learning tight-binding approach for large-scale electronic simulations at finite temperatures with ab initio accuracy}.
\bjtitle{Nat. Commun.}
\bvolume{15}(\bissue{1}),
\bfpage{6772}
(\byear{2024})
\doiurl{10.1038/s41467-024-51006-4}
\end{barticle}
\endbibitem

\bibitem[\protect\citeauthoryear{Qiao et~al.}{2022}]{bib19}
\begin{barticle}
\bauthor{\bsnm{Qiao}, \binits{Z.}},
\bauthor{\bsnm{Christensen}, \binits{A.S.}},
\bauthor{\bsnm{Welborn}, \binits{M.}},
\bauthor{\bsnm{Manby}, \binits{F.R.}},
\bauthor{\bsnm{Anandkumar}, \binits{A.}},
\bauthor{\bsnm{Miller}, \binits{r.} \bsuffix{T.~F.}}:
\batitle{Informing geometric deep learning with electronic interactions to accelerate quantum chemistry}.
\bjtitle{Proc. Natl. Acad. Sci. USA}
\bvolume{119}(\bissue{31}),
\bfpage{2205221119}
(\byear{2022})
\doiurl{10.1073/pnas.2205221119}
\end{barticle}
\endbibitem

\bibitem[\protect\citeauthoryear{Zhong et~al.}{2024}]{bib20}
\begin{barticle}
\bauthor{\bsnm{Zhong}, \binits{Y.}},
\bauthor{\bsnm{Yu}, \binits{H.}},
\bauthor{\bsnm{Yang}, \binits{J.}},
\bauthor{\bsnm{Guo}, \binits{X.}},
\bauthor{\bsnm{Xiang}, \binits{H.}},
\bauthor{\bsnm{Gong}, \binits{X.}}:
\batitle{Universal machine learning kohn-sham hamiltonian for materials}.
\bjtitle{Chin. Phys. Lett.}
\bvolume{41},
\bfpage{077103}
(\byear{2024})
\doiurl{10.1088/0256-307x/41/7/077103}
\end{barticle}
\endbibitem

\bibitem[\protect\citeauthoryear{Grisafi et~al.}{2018}]{bib21}
\begin{barticle}
\bauthor{\bsnm{Grisafi}, \binits{A.}},
\bauthor{\bsnm{Wilkins}, \binits{D.M.}},
\bauthor{\bsnm{Csanyi}, \binits{G.}},
\bauthor{\bsnm{Ceriotti}, \binits{M.}}:
\batitle{Symmetry-adapted machine learning for tensorial properties of atomistic systems}.
\bjtitle{Phys. Rev. Lett.}
\bvolume{120}(\bissue{3}),
\bfpage{036002}
(\byear{2018})
\doiurl{10.1103/PhysRevLett.120.036002}
\end{barticle}
\endbibitem

\bibitem[\protect\citeauthoryear{Morrison and Parker}{1987}]{bib22}
\begin{barticle}
\bauthor{\bsnm{Morrison}, \binits{M.A.}},
\bauthor{\bsnm{Parker}, \binits{G.A.}}:
\batitle{A guide to rotations in quantum-mechanics}.
\bjtitle{Aust. J. Phys.}
\bvolume{40}(\bissue{4}),
\bfpage{465}--\blpage{497}
(\byear{1987})
\end{barticle}
\endbibitem

\bibitem[\protect\citeauthoryear{Jones and Albers}{2009}]{bib23}
\begin{barticle}
\bauthor{\bsnm{Jones}, \binits{M.D.}},
\bauthor{\bsnm{Albers}, \binits{R.C.}}:
\batitle{Spin-orbit coupling in an f-electron tight-binding model: Electronic properties of th, u, and pu}.
\bjtitle{Phys. Rev. B}
\bvolume{79}(\bissue{4}),
\bfpage{045107}
(\byear{2009})
\doiurl{10.1103/PhysRevB.79.045107}
\end{barticle}
\endbibitem

\bibitem[\protect\citeauthoryear{Fernández-Seivane et~al.}{2006}]{RN2214}
\begin{barticle}
\bauthor{\bsnm{Fernández-Seivane}, \binits{L.}},
\bauthor{\bsnm{Oliveira}, \binits{M.A.}},
\bauthor{\bsnm{Sanvito}, \binits{S.}},
\bauthor{\bsnm{Ferrer}, \binits{J.}}:
\batitle{J. phys.: Condens. matter}.
\bjtitle{Journal of Physics: Condensed Matter}
\bvolume{18}(\bissue{34}),
\bfpage{7999}--\blpage{8013}
(\byear{2006})
\doiurl{10.1088/0953-8984/18/34/012}
\end{barticle}
\endbibitem

\bibitem[\protect\citeauthoryear{Kurita and Koretsune}{2020}]{RN2215}
\begin{botherref}
\oauthor{\bsnm{Kurita}, \binits{K.}},
\oauthor{\bsnm{Koretsune}, \binits{T.}}:
Systematic first-principles study of the on-site spin-orbit coupling in crystals.
Phys. Rev. B
\textbf{102}(4)
(2020)
\doiurl{10.1103/PhysRevB.102.045109}
\end{botherref}
\endbibitem

\bibitem[\protect\citeauthoryear{Cuadrado et~al.}{2021}]{RN2216}
\begin{botherref}
\oauthor{\bsnm{Cuadrado}, \binits{R.}},
\oauthor{\bsnm{Robles}, \binits{R.}},
\oauthor{\bsnm{García}, \binits{A.}},
\oauthor{\bsnm{Pruneda}, \binits{M.}},
\oauthor{\bsnm{Ordejón}, \binits{P.}},
\oauthor{\bsnm{Ferrer}, \binits{J.}},
\oauthor{\bsnm{Cerdá}, \binits{J.I.}}:
Validity of the on-site spin-orbit coupling approximation.
Phys. Rev. B
\textbf{104}(19)
(2021)
\doiurl{10.1103/PhysRevB.104.195104}
\end{botherref}
\endbibitem

\bibitem[\protect\citeauthoryear{Batatia et~al.}{2022}]{bib25}
\begin{botherref}
\oauthor{\bsnm{Batatia}, \binits{I.}},
\oauthor{\bsnm{Kovács}, \binits{D.P.}},
\oauthor{\bsnm{Simm}, \binits{G.N.C.}},
\oauthor{\bsnm{Ortner}, \binits{C.}},
\oauthor{\bsnm{Csányi}, \binits{G.}}:
MACE: higher order equivariant message passing neural networks for fast and accurate force fields.
Curran Associates Inc.
(2022)
\end{botherref}
\endbibitem

\bibitem[\protect\citeauthoryear{Jain et~al.}{2013}]{bib26}
\begin{barticle}
\bauthor{\bsnm{Jain}, \binits{A.}},
\bauthor{\bsnm{Ong}, \binits{S.P.}},
\bauthor{\bsnm{Hautier}, \binits{G.}},
\bauthor{\bsnm{Chen}, \binits{W.}},
\bauthor{\bsnm{Richards}, \binits{W.D.}},
\bauthor{\bsnm{Dacek}, \binits{S.}},
\bauthor{\bsnm{Cholia}, \binits{S.}},
\bauthor{\bsnm{Gunter}, \binits{D.}},
\bauthor{\bsnm{Skinner}, \binits{D.}},
\bauthor{\bsnm{Ceder}, \binits{G.}},
\bauthor{\bsnm{Persson}, \binits{K.A.}}:
\batitle{Commentary: The materials project: A materials genome approach to accelerating materials innovation}.
\bjtitle{APL Mater.}
\bvolume{1}(\bissue{1}),
\bfpage{011002}
(\byear{2013})
\doiurl{10.1063/1.4812323}
\end{barticle}
\endbibitem

\bibitem[\protect\citeauthoryear{Hasan and Kane}{2010}]{bib27}
\begin{barticle}
\bauthor{\bsnm{Hasan}, \binits{M.Z.}},
\bauthor{\bsnm{Kane}, \binits{C.L.}}:
\batitle{Colloquium: Topological insulators}.
\bjtitle{Rev. Mod. Phys}
\bvolume{82}(\bissue{4}),
\bfpage{3045}--\blpage{3067}
(\byear{2010})
\doiurl{10.1103/RevModPhys.82.3045}
\end{barticle}
\endbibitem

\bibitem[\protect\citeauthoryear{Fukui et~al.}{2005}]{bib28}
\begin{barticle}
\bauthor{\bsnm{Fukui}, \binits{T.}},
\bauthor{\bsnm{Hatsugai}, \binits{Y.}},
\bauthor{\bsnm{Suzuki}, \binits{H.}}:
\batitle{Chern numbers in discretized brillouin zone: Efficient method of computing (spin) hall conductances}.
\bjtitle{J. Phys. Soc. Jpn.}
\bvolume{74}(\bissue{6}),
\bfpage{1674}--\blpage{1677}
(\byear{2005})
\doiurl{10.1143/jpsj.74.1674}
\end{barticle}
\endbibitem

\bibitem[\protect\citeauthoryear{Wu et~al.}{2018}]{bib29}
\begin{barticle}
\bauthor{\bsnm{Wu}, \binits{Q.S.}},
\bauthor{\bsnm{Zhang}, \binits{S.N.}},
\bauthor{\bsnm{Song}, \binits{H.F.}},
\bauthor{\bsnm{Troyer}, \binits{M.}},
\bauthor{\bsnm{Soluyanov}, \binits{A.A.}}:
\batitle{Wanniertools: An open-source software package for novel topological materials}.
\bjtitle{Comput. Phys. Commun.}
\bvolume{224},
\bfpage{405}--\blpage{416}
(\byear{2018})
\doiurl{10.1016/j.cpc.2017.09.033}
\end{barticle}
\endbibitem

\bibitem[\protect\citeauthoryear{Marzari et~al.}{2012}]{bib30}
\begin{barticle}
\bauthor{\bsnm{Marzari}, \binits{N.}},
\bauthor{\bsnm{Mostofi}, \binits{A.A.}},
\bauthor{\bsnm{Yates}, \binits{J.R.}},
\bauthor{\bsnm{Souza}, \binits{I.}},
\bauthor{\bsnm{Vanderbilt}, \binits{D.}}:
\batitle{Maximally localized wannier functions: Theory and applications}.
\bjtitle{Rev. Mod. Phys.}
\bvolume{84}(\bissue{4}),
\bfpage{1419}--\blpage{1475}
(\byear{2012})
\doiurl{10.1103/RevModPhys.84.1419}
\end{barticle}
\endbibitem

\bibitem[\protect\citeauthoryear{Merchant et~al.}{2023}]{bib31}
\begin{barticle}
\bauthor{\bsnm{Merchant}, \binits{A.}},
\bauthor{\bsnm{Batzner}, \binits{S.}},
\bauthor{\bsnm{Schoenholz}, \binits{S.S.}},
\bauthor{\bsnm{Aykol}, \binits{M.}},
\bauthor{\bsnm{Cheon}, \binits{G.}},
\bauthor{\bsnm{Cubuk}, \binits{E.D.}}:
\batitle{Scaling deep learning for materials discovery}.
\bjtitle{Nature}
\bvolume{624}(\bissue{7990}),
\bfpage{80}--\blpage{85}
(\byear{2023})
\doiurl{10.1038/s41586-023-06735-9}
\end{barticle}
\endbibitem

\bibitem[\protect\citeauthoryear{Haastrup et~al.}{2018}]{bib32}
\begin{barticle}
\bauthor{\bsnm{Haastrup}, \binits{S.}},
\bauthor{\bsnm{Strange}, \binits{M.}},
\bauthor{\bsnm{Pandey}, \binits{M.}},
\bauthor{\bsnm{Deilmann}, \binits{T.}},
\bauthor{\bsnm{Schmidt}, \binits{P.S.}},
\bauthor{\bsnm{Hinsche}, \binits{N.F.}},
\bauthor{\bsnm{Gjerding}, \binits{M.N.}},
\bauthor{\bsnm{Torelli}, \binits{D.}},
\bauthor{\bsnm{Larsen}, \binits{P.M.}},
\bauthor{\bsnm{Riis-Jensen}, \binits{A.C.}},
\bauthor{\bsnm{Gath}, \binits{J.}},
\bauthor{\bsnm{Jacobsen}, \binits{K.W.}},
\bauthor{\bsnm{Mortensen}, \binits{J.J.}},
\bauthor{\bsnm{Olsen}, \binits{T.}},
\bauthor{\bsnm{Thygesen}, \binits{K.S.}}:
\batitle{The computational 2d materials database: high-throughput modeling and discovery of atomically thin crystals}.
\bjtitle{2d Mater.}
\bvolume{5}(\bissue{4}),
\bfpage{042002}
(\byear{2018})
\doiurl{10.1088/2053-1583/aacfc1}
\end{barticle}
\endbibitem

\bibitem[\protect\citeauthoryear{Gjerding et~al.}{2021}]{bib33}
\begin{barticle}
\bauthor{\bsnm{Gjerding}, \binits{M.N.}},
\bauthor{\bsnm{Taghizadeh}, \binits{A.}},
\bauthor{\bsnm{Rasmussen}, \binits{A.}},
\bauthor{\bsnm{Ali}, \binits{S.}},
\bauthor{\bsnm{Bertoldo}, \binits{F.}},
\bauthor{\bsnm{Deilmann}, \binits{T.}},
\bauthor{\bsnm{Knosgaard}, \binits{N.R.}},
\bauthor{\bsnm{Kruse}, \binits{M.}},
\bauthor{\bsnm{Larsen}, \binits{A.H.}},
\bauthor{\bsnm{Manti}, \binits{S.}},
\bauthor{\bsnm{Pedersen}, \binits{T.G.}},
\bauthor{\bsnm{Petralanda}, \binits{U.}},
\bauthor{\bsnm{Skovhus}, \binits{T.}},
\bauthor{\bsnm{Svendsen}, \binits{M.K.}},
\bauthor{\bsnm{Mortensen}, \binits{J.J.}},
\bauthor{\bsnm{Olsen}, \binits{T.}},
\bauthor{\bsnm{Thygesen}, \binits{K.S.}}:
\batitle{Recent progress of the computational 2d materials database (c2db)}.
\bjtitle{2d Mater.}
\bvolume{8}(\bissue{4}),
\bfpage{044002}
(\byear{2021})
\doiurl{10.1088/2053-1583/ac1059}
\end{barticle}
\endbibitem

\bibitem[\protect\citeauthoryear{Unke and Meuwly}{2019}]{bib34}
\begin{barticle}
\bauthor{\bsnm{Unke}, \binits{O.T.}},
\bauthor{\bsnm{Meuwly}, \binits{M.}}:
\batitle{Physnet: A neural network for predicting energies, forces, dipole moments, and partial charges}.
\bjtitle{J. Chem. Theory Comput.}
\bvolume{15}(\bissue{6}),
\bfpage{3678}--\blpage{3693}
(\byear{2019})
\doiurl{10.1021/acs.jctc.9b00181}
\end{barticle}
\endbibitem

\bibitem[\protect\citeauthoryear{Thomas et~al.}{2018}]{bib35}
\begin{botherref}
\oauthor{\bsnm{Thomas}, \binits{N.}},
\oauthor{\bsnm{Smidt}, \binits{T.}},
\oauthor{\bsnm{Kearnes}, \binits{S.}},
\oauthor{\bsnm{Yang}, \binits{L.}},
\oauthor{\bsnm{Li}, \binits{L.}},
\oauthor{\bsnm{Kohlhoff}, \binits{K.}},
\oauthor{\bsnm{Riley}, \binits{P.}}:
Tensor field networks: Rotation- and translation-equivariant neural networks for 3d point clouds.
Preprint at https://arxiv.org/abs/1802.08219v3
(2018)
\end{botherref}
\endbibitem

\bibitem[\protect\citeauthoryear{Batzner et~al.}{2022}]{bib36}
\begin{barticle}
\bauthor{\bsnm{Batzner}, \binits{S.}},
\bauthor{\bsnm{Musaelian}, \binits{A.}},
\bauthor{\bsnm{Sun}, \binits{L.X.}},
\bauthor{\bsnm{Geiger}, \binits{M.}},
\bauthor{\bsnm{Mailoa}, \binits{J.P.}},
\bauthor{\bsnm{Kornbluth}, \binits{M.}},
\bauthor{\bsnm{Molinari}, \binits{N.}},
\bauthor{\bsnm{Smidt}, \binits{T.E.}},
\bauthor{\bsnm{Kozinsky}, \binits{B.}}:
\batitle{E(3)-equivariant graph neural networks for data-efficient and accurate interatomic potentials}.
\bjtitle{Nat. Commun.}
\bvolume{13}(\bissue{1}),
\bfpage{2453}
(\byear{2022})
\doiurl{10.1038/s41467-022-29939-5}
\end{barticle}
\endbibitem

\bibitem[\protect\citeauthoryear{Loshchilov and Hutter}{2017}]{bib37}
\begin{botherref}
\oauthor{\bsnm{Loshchilov}, \binits{I.}},
\oauthor{\bsnm{Hutter}, \binits{F.}}:
Decoupled weight decay regularization.
Preprint at https://doi.org/10.48550/arXiv.1711.05101
(2017)
\end{botherref}
\endbibitem

\bibitem[\protect\citeauthoryear{Reddi et~al.}{2019}]{bib38}
\begin{botherref}
\oauthor{\bsnm{Reddi}, \binits{S.J.}},
\oauthor{\bsnm{Kale}, \binits{S.}},
\oauthor{\bsnm{Kumar}, \binits{S.}}:
On the Convergence of Adam and Beyond
(2019).
\url{https://ui.adsabs.harvard.edu/abs/2019arXiv190409237R}
\end{botherref}
\endbibitem

\bibitem[\protect\citeauthoryear{Ozaki}{2003}]{bib39}
\begin{barticle}
\bauthor{\bsnm{Ozaki}, \binits{T.}}:
\batitle{Variationally optimized atomic orbitals for large-scale electronic structures}.
\bjtitle{Phys. Rev. B}
\bvolume{67}(\bissue{15}),
\bfpage{155108}
(\byear{2003})
\doiurl{10.1103/PhysRevB.67.155108}
\end{barticle}
\endbibitem

\end{thebibliography}

\end{document}